\documentclass[preprint]{ptephy_om}

\preprintnumber{KYUSHU-HET-387} 

\usepackage{graphics}

\usepackage{amsmath} 
\usepackage{amsthm} 
\usepackage{url} 
\usepackage{stmaryrd}
\usepackage{tikz}
\usepackage{tikz-feynhand}
\usepackage{mathtools}
\usepackage{bm}
\usepackage{booktabs}
\usepackage{subcaption}
\usepackage{color}

\DeclareMathOperator{\tr}{tr}

\newcommand{\Slash}[1]{{\ooalign{\hfil/\hfil\crcr$#1$}}}
\numberwithin{equation}{section}

\allowdisplaybreaks

\begin{document}

\title{Wilsonian exact renormalization group flow in QED with the Pauli term}


\author[1]{Sorato Nagao}
\affil[1]{Department of Physics, Kyushu University, 744 Motooka, Nishi-ku,
Fukuoka 819-0395, Japan}

\author[1,2]{Hiroshi Suzuki}
\affil[2]{Quantum and Spacetime Research Institute (QuaSR), Kyushu University,
744 Motooka, Nishi-ku, Fukuoka 819-0395, Japan}





\begin{abstract}%
We study the renormalization group (RG) flow of quantum electrodynamics (QED)
with a Pauli term, $\kappa_\tau\Bar{\psi}\sigma_{\mu\nu}F_{\mu\nu}\psi$, by using
the gradient flow exact renormalization group (GFERG). In the GFERG, the gauge
Ward--Takahashi (WT) identity takes the conventional form and is exactly
preserved, and the gauge coupling flows only through the photon anomalous
dimension~$\gamma_A$. Consequently, at any fixed point with a vanishing gauge
coupling, the eigenvalue in the direction of the gauge coupling is
exactly~$-\gamma_A^*$, and the gauge coupling is relevant if and only if the
vacuum polarization makes~$\gamma_A^*$ negative. In the leading large-$N_f$
approximation, the vacuum polarization induced by the Pauli coupling yields
such a fixed point, with~$\gamma_A^*=-1$ and three relevant directions. At
finite~$N_f$, however, the $N_f$-independent one-loop effects on the fermion
line eliminate it unless $N_f$ is at least several tens. For~$N_f=1$, a search
over the whole plane of the Pauli coupling and the mass parameter in the
subspace~$e_\tau=0$, where our flow equation is exactly of one-loop order,
finds a single fixed point, at strong Pauli coupling and a finite mass
parameter. Like one of the fixed points of the asymptotically safe QED proposed
by Gies and Ziebell, it has a finite mass parameter and two relevant
directions. At their fixed point, the gauge coupling is relevant. At ours, in
contrast, $\gamma_A^*>0$, so that the gauge coupling is irrelevant, and the
fixed point cannot provide a UV completion of QED with a nonzero charge. Along
the full flow for~$N_f=1$ on the critical surface connected to the Gaussian
fixed point, $\kappa_\tau$ is irrelevant and is attracted to an induced value
proportional to~$e_\tau^3$ at weak coupling, while $e_\tau^2$ is marginally
irrelevant as in QED without the Pauli term.
\end{abstract}

\subjectindex{B02,B05,B31,B32}

\maketitle

\section{Introduction}
\label{sec:1}
Whether four-dimensional quantum electrodynamics (QED) possesses a nontrivial
continuum limit is a long-standing question. Perturbatively, the gauge coupling
grows toward the ultraviolet (UV) and hits the Landau pole~\cite{Landau:1955},
and lattice simulations of non-compact QED suggest that the theory is
trivial~\cite{Gockeler:1997dn,Kim:2000rr,Kim:2001am}. One possible way out
within QED itself is a strongly coupled fixed point associated with dynamical
chiral symmetry breaking~\cite{Miransky:1984ef,Bardeen:1985sm,Leung:1985sn}. In
a previous paper~\cite{Nagao:2026uxd}, we studied the gauged
Nambu--Jona-Lasinio model by the gradient flow exact renormalization group
(GFERG) and found that, within our truncation, the merging point of the
Gaussian-type and NJL-type quasi-fixed points is not a true fixed point,
because the photon anomalous dimension~$\gamma_A$ stays positive there.

A different possibility was proposed by Gies and Ziebell~\cite{Gies:2020xuh}.
In their functional RG analysis, QED is extended by the Pauli
term~$\Bar{\psi}\sigma_{\mu\nu}F_{\mu\nu}\psi$, which is perturbatively
nonrenormalizable. For~$N_f=1$, they found two non-Gaussian fixed points, both
with a vanishing gauge coupling and a finite Pauli coupling: one at a vanishing
mass parameter with three relevant directions, which we denote by~$\mathcal{C}$
following~Ref.~\cite{Gies:2020xuh}, and one at a finite dimensionless mass
parameter with two relevant directions, $\mathcal{B}$. At both of them, the
fermion anomalous dimension is large and negative, and the gauge coupling is
relevant, i.e., asymptotically free; the fixed points may thus provide a UV
completion of QED. (The fixed point~$\mathcal{B}$ predicts an anomalous
magnetic moment of the electron incompatible with observation, so that physical
QED was argued to be in the universality class
of~$\mathcal{C}$~\cite{Gies:2020xuh}.) The fixed-point structure was further
studied in general dimensions~\cite{Gies:2022aiz} and including chiral
four-fermion interactions~\cite{Gies:2024ugz}. The idea that higher-order terms
might cure the Landau pole has also been discussed from the
effective-field-theory viewpoint~\cite{Djukanovic:2017thn}.

These fixed points lie at strong Pauli coupling, where the fermion anomalous
dimension is large and the reliability of a truncation is not obvious.
Moreover, in the functional RG, the regulator breaks the gauge symmetry, which
is then retained only through a modified Ward--Takahashi (WT) identity; it is
not easy to respect this identity within a truncation. It is therefore
important to examine these fixed points in a framework in which the gauge
symmetry is manifestly preserved and the approximation can be improved
systematically.

The GFERG~\cite{Sonoda:2020vut,Sonoda:2022fmk,Sonoda:2025tyu} (see also
Refs.~\cite{Sonoda:2019ibh,Miyakawa:2021hcx,Miyakawa:2021wus,Abe:2022smm,Miyakawa:2022qbz,Miyakawa:2023yob,Haruna:2023spq,Nagao:2025sri}) is suited to this
purpose. In the GFERG, the coarse-graining is defined by the gradient
flow~\cite{Narayanan:2006rf,Luscher:2010iy,Luscher:2011bx,Luscher:2013cpa}, and
the 1PI Wilson action of the $U(1)$ gauge theory is the sum of a manifestly
gauge-invariant functional and the standard gauge-fixing
term~\cite{Sonoda:2022fmk}. In Refs.~\cite{Nagao:2026vaf,Nagao:2026gbq}, we
derived and solved the GFERG equation for a gauge-invariant ansatz for QED;
in~Ref.~\cite{Nagao:2026uxd}, the ansatz was extended by the four-fermion
interactions.

In this paper, we add the Pauli term
$\kappa_\tau\Bar{\psi}\sigma_{\mu\nu}F_{\mu\nu}\psi$ to the QED ansatz
of~Refs.~\cite{Nagao:2026vaf,Nagao:2026gbq}, where $\kappa_\tau$ is the Pauli
coupling. Here and in what follows, $\tau$ parametrizes the scale
logarithmically, with~$\tau\to\infty$ the infrared (IR) direction, and the
couplings, such as~$\kappa_\tau$, the gauge coupling~$e_\tau$, and the mass
parameter~$m_\tau$, are made dimensionless by the cutoff scale at~$\tau$. The
ansatz also contains the kinetic function of the gauge field,
$\mathcal{K}_\tau(k^2)$, normalized as~$\mathcal{K}_\tau(0)=1$; in the
fixed-point analysis, we either set~$\mathcal{K}_\tau=1$ or use the stationary
solution of its flow equation. The new coupling~$\kappa_\tau$ requires the
magnetic part of the fermion--photon three-point function on the right-hand
side of the GFERG equation; we evaluate this by extending the numerical method
of~Ref.~\cite{Nagao:2026uxd}.

Our findings are as follows. (i)~Since the GFERG preserves the WT identity
exactly, the gauge coupling~$e_\tau$ flows only through the photon anomalous
dimension~$\gamma_A$. Therefore, at any fixed point with~$e_*=0$, the
eigenvalue in the direction of the gauge coupling is exactly~$-\gamma_A^*$,
irrespective of the truncation of the fermion sector. The gauge coupling is
relevant if and only if the vacuum polarization makes~$\gamma_A^*$ negative,
and the fermion anomalous dimension does not enter. (ii)~In the leading
large-$N_f$ approximation, the vacuum polarization caused by the Pauli coupling
gives a fixed point with~$e_*=0$, $\kappa_*\neq0$, and~$\gamma_A^*=-1$
in~$D=4$, with three relevant directions. Its number of relevant directions is
the same as that of the fixed point~$\mathcal{C}$ of~Ref.~\cite{Gies:2020xuh},
but the mechanism is different: here the vacuum polarization makes both the
gauge and the Pauli couplings relevant, whereas in~Ref.~\cite{Gies:2020xuh} the
photon anomalous dimension vanishes and the fermion anomalous dimension is
large and negative. (iii)~All the two- and three-loop terms of the GFERG
equation are proportional to the gauge coupling~$e_\tau$. Therefore, in the
subspace~$e_\tau=0$, which is preserved by the flow, the right-hand side of the
GFERG equation is exactly of one-loop order. The vacuum polarization is
proportional to~$N_f\kappa^2$, whereas the one-loop effects on the fermion
line---the fermion anomalous dimension and, in particular, the additive mass
renormalization---are proportional to~$\kappa^2$ with coefficients independent
of~$N_f$, e.g., $2\gamma_\psi\simeq0.061\kappa^2$ and the critical
mass~$m_c\simeq0.04\kappa^2$. In the large-$N_f$ approximation, in
which~$N_f\kappa^2$ is kept fixed, the latter are suppressed by~$1/N_f$ and are
neglected. At finite~$N_f$, however, they are not negligible: at the
large-$N_f$ fixed point,
$\kappa_*^2=\Tilde{\kappa}_*^2/(\tr(1)N_f)\simeq77/N_f$, where
$\Tilde{\kappa}:=\kappa_\tau\sqrt{\tr(1)N_f}$ is the Pauli coupling rescaled for
the large-$N_f$ limit, so that $2\gamma_\psi\simeq4.7/N_f$ and~$m_c\simeq3/N_f$,
which are not small unless~$N_f$ is large. As a result, a fixed point at weak
Pauli coupling, if any, requires at least several tens of flavors (it appears
for~$N_f\gtrsim56$ with~$\mathcal{K}_\tau=1$, and is not found up to~$N_f=64$
with the stationary~$\mathcal{K}_\tau$), and certainly does not exist
for~$N_f=1$. (iv)~For~$N_f=1$, we search the whole plane of the Pauli coupling
and the mass parameter in the subspace~$e_\tau=0$, and find a single fixed
point, at strong Pauli coupling and a finite mass parameter; the latter is the
value of the critical mass there, so that the fixed point lies on the critical
surface. Like the fixed point~$\mathcal{B}$ of~Ref.~\cite{Gies:2020xuh}, it
has a large negative fermion anomalous dimension and two relevant directions.
At~$\mathcal{B}$, the gauge coupling is relevant. At our fixed point, in
contrast, $\gamma_A^*>0$, so that the two relevant directions are those of the
Pauli coupling and the mass, and the gauge coupling is irrelevant. All the
trajectories emanating from this fixed point stay in the subspace~$e_\tau=0$,
and it does not provide a UV completion of QED with a nonzero charge. Since the
fermion anomalous dimension exceeds unity in magnitude at this fixed point, its
existence lies outside the range of validity of our truncation; the irrelevance
of the gauge coupling, however, requires only~$\gamma_A^*>0$, which holds with
a large margin. (v)~For~$N_f=1$, we follow the full flow with~$e_\tau\neq0$ on
the critical surface connected to the Gaussian fixed point~$m=0$. The Pauli
coupling is irrelevant at the Gaussian fixed point. Since $\kappa_\tau$ relaxes
as~$e^{-\tau}$ while $e_\tau$ runs only logarithmically, $\kappa_\tau$ is
attracted to the value~$\kappa_{\text{ind}}(e_\tau)$ at which
$d\kappa_\tau/d\tau\simeq0$ for the current~$e_\tau$. This induced Pauli
coupling is proportional to~$e_\tau^3$ at weak coupling and comes from the
one-loop vertex correction that flips the chirality. It is analogous to the
anomalous magnetic moment of the electron~\cite{Schwinger:1948iu}, for which
the chirality flip is provided by the fermion mass; here, it is provided by the
chirality breaking of the regularization, so that~$\kappa_{\text{ind}}$ is
generated even at~$m=0$. The gauge coupling is marginally irrelevant as in QED
without the Pauli term, and we find no fixed point with~$e_\tau\neq0$ on the
critical surface connected to~$m=0$ in the region we studied. Unlike in the
subspace~$e_\tau=0$, however, the critical surface at strong Pauli coupling and
its other branches are not examined for~$e_\tau\neq0$.

This paper is organized as follows. In~Sect.~\ref{sec:2}, we recapitulate the
GFERG flow equation and introduce the ansatz for the 1PI Wilson action. We then
derive the general form of the flow equations for the ansatz.
Section~\ref{sec:3} treats the leading large-$N_f$ approximation,
Sect.~\ref{sec:4} the exact one-loop flow in the subspace~$e_\tau=0$ for
general~$N_f$, including the fixed point at a finite mass parameter for~$N_f=1$
and the comparison with the functional RG analysis~\cite{Gies:2020xuh}, and
Sect.~\ref{sec:5} the full flow for~$N_f=1$ on the critical surface connected
to~$m=0$ in~$D=4$. Section~\ref{sec:6} is the conclusion. The numerical method
is described briefly in~Sect.~\ref{sec:5.1}.

\section{Formulation}
\label{sec:2}
The GFERG flow equation for the 1PI Wilson
action~$\mathit{\Gamma}_\tau[\mathcal{A},\Bar\Psi,\Psi]$ of the $U(1)$ gauge
theory is given by\footnote{%
$D$ denotes the spacetime dimension, $\epsilon:=(4-D)/2$,
and~$\int_p:=\int d^Dp/(2\pi)^D$.}
\begin{align}
   &\frac{\partial}{\partial\tau}
   \mathit{\Gamma}_\tau[\mathcal{A},\Bar{\Psi},\Psi]
\notag\\
   &=\int_p\,
   \biggl[
   \left(2p^2+\frac{D+2}{2}-\epsilon-\gamma_A
   +p\cdot\frac{\partial}{\partial p}\right)\mathcal{A}_\mu(p)
   \cdot\frac{\delta\mathit{\Gamma}_\tau}{\delta\mathcal{A}_\mu(p)}
\notag\\
   &\qquad\qquad{}
   +\left(
   2p^2+\frac{D+1}{2}-\gamma_\psi+p\cdot\frac{\partial}{\partial p}
   \right)
   \Bar{\Psi}(p)\cdot
   \frac{\delta}{\delta\Bar{\Psi}(p)}\mathit{\Gamma}_\tau
\notag\\
   &\qquad\qquad{}
   +\left(
   2p^2+\frac{D+1}{2}-\gamma_\psi+p\cdot\frac{\partial}{\partial p}
   \right)
   \Psi(p)\cdot
   \frac{\delta}{\delta\Psi(p)}\mathit{\Gamma}_\tau\biggr]
\notag\\
   &\qquad{}
   +\int_p\,
   \biggl[
   \left(2p^2-\epsilon-\gamma_A+1\right)
   \frac{\delta\mathcal{A}_\mu(p)}{\delta A_\mu(p)}
\notag\\
   &\qquad\qquad\qquad{}
   +\left(-2p^2+\gamma_\psi-\frac{1}{2}\right)
   \frac{\delta}{\delta\Bar{\psi}(p)}
   \Bar{\Psi}(p)
   +\left(-2p^2+\gamma_\psi-\frac{1}{2}\right)
   \frac{\delta}{\delta\psi(p)}
   \Psi(p)
   \biggr]
\notag\\
   &\qquad{}
   +\int_{p,k}\,
   (-4)e_\tau p_\mu\left[
   \frac{\delta}{\delta\Bar{\psi}(p+k)}
   +\frac{\delta}{\delta\Bar{\Psi}(p+k)}\mathit{\Gamma}_\tau
   \right]
   \left[
   \mathcal{A}_\mu(k)
   +\frac{\delta}{\delta A_\mu(-k)}\right]
   \Bar{\Psi}(p)
\notag\\
   &\qquad{}
   +\int_{p,k,l}\,
   (-2)e_\tau^2\left[
   \frac{\delta}{\delta\Bar{\psi}(p+k+l)}
   +\frac{\delta}{\delta\Bar{\Psi}(p+k+l)}\mathit{\Gamma}_\tau
   \right]
   \left[
   \mathcal{A}_\mu(k)
   +\frac{\delta}{\delta A_\mu(-k)}\right]
\notag\\
   &\qquad\qquad\qquad\qquad\qquad\qquad{}
   \times\left[
   \mathcal{A}_\mu(l)
   +\frac{\delta}{\delta A_\mu(-l)}\right]
   \Bar{\Psi}(p)
\notag\\
   &\qquad{}
   +\int_{p,k}\,
   4e_\tau p_\mu\left[
   \frac{\delta}{\delta\psi(p+k)}
   +\frac{\delta}{\delta\Psi(p+k)}\mathit{\Gamma}_\tau
   \right]
   \left[
   \mathcal{A}_\mu(k)
   +\frac{\delta}{\delta A_\mu(-k)}\right]
   \Psi(p)
\notag\\
   &\qquad{}
   +\int_{p,k,l}\,
   (-2)e_\tau^2\left[
   \frac{\delta}{\delta\psi(p+k+l)}
   +\frac{\delta}{\delta\Psi(p+k+l)}\mathit{\Gamma}_\tau
   \right]
   \left[
   \mathcal{A}_\mu(k)
   +\frac{\delta}{\delta A_\mu(-k)}\right]
\notag\\
   &\qquad\qquad\qquad\qquad\qquad\qquad{}
   \times\left[
   \mathcal{A}_\mu(l)
   +\frac{\delta}{\delta A_\mu(-l)}\right]
   \Psi(p),
\label{eq:(2.1)}
\end{align}
where the parameter~$\tau$ parametrizes the scale logarithmically, and
$\tau\to\infty$ is the IR direction; $\gamma_A$ and~$\gamma_\psi$ are the
anomalous dimensions of the gauge and the fermion fields, respectively. The
variables $\{\mathcal{A}_\mu,\Bar{\Psi},\Psi\}$ are the arguments of the 1PI
Wilson action, whereas $\{A_\mu,\Bar{\psi},\psi\}$ are those of the (non-1PI)
Wilson action; these two sets are related by the Legendre transformation
\begin{equation}
   A_\mu(x)=\mathcal{A}_\mu(x)
   -\frac{\delta\mathit{\Gamma}_\tau}{\delta\mathcal{A}_\mu(x)},\qquad
   \psi(x)=\Psi(x)-i\frac{\delta}{\delta\Bar{\Psi}(x)}\mathit{\Gamma}_\tau,
   \qquad
   \Bar{\psi}(x)
   =\Bar{\Psi}(x)+i\frac{\delta}{\delta\Psi(x)}\mathit{\Gamma}_\tau.
\label{eq:(2.2)}
\end{equation}
Field-independent constants (which contribute to the vacuum energy) are
omitted in~Eq.~\eqref{eq:(2.1)}, which is equivalent to Eq.~(2.3)
of~Ref.~\cite{Nagao:2026uxd}. In conjunction with this, the RG flow equations
for the gauge coupling~$e_\tau$ and the gauge-fixing parameter~$\xi_\tau$ are
given by
\begin{equation}
   \frac{d}{d\tau}e_\tau^2=-2\gamma_Ae_\tau^2,\qquad
   \frac{d}{d\tau}\xi_\tau=2(\epsilon+\gamma_A)\xi_\tau.
\label{eq:(2.3)}
\end{equation}
In what follows, we work in the Landau gauge, $\xi_\tau=0$, which is preserved
under~Eq.~\eqref{eq:(2.3)}.

The first relation in~Eq.~\eqref{eq:(2.3)} is a consequence of the WT
identity~\cite{Nagao:2026vaf}: the gauge coupling appears only through the
covariant derivative, and is renormalized only through the normalization of
the gauge field. This has a simple but important consequence for fixed points
with a vanishing gauge coupling, which we discuss in~Sects.~\ref{sec:3}
and~\ref{sec:4}. Since $de_\tau^2/d\tau\propto e_\tau^2$, the subspace~$e_\tau=0$
is preserved, and the stability matrix at a fixed point with~$e_*=0$ is block
triangular. The eigenvalue in the direction of~$e_\tau$ is therefore exactly
\begin{equation}
   \lambda_e=-\gamma_A^*,
\label{eq:(2.4)}
\end{equation}
where $\gamma_A^*$ is the photon anomalous dimension at the fixed point. The
gauge coupling is relevant at such a fixed point if and only if~$\gamma_A^*<0$,
i.e., if the vacuum polarization dominates; the fermion anomalous
dimension~$\gamma_\psi$ does not enter~$\lambda_e$, whatever the ansatz for the
fermion sector is.

A step-by-step derivation of the flow equations for the $U(1)$ theory,
Eqs.~\eqref{eq:(2.1)} and~\eqref{eq:(2.3)}, is given
in~Ref.~\cite{Nagao:2026vaf}; see also Ref.~\cite{Nagao:2026uxd}. On the
right-hand side of~Eq.~\eqref{eq:(2.1)}, the first three lines are the linear
terms associated with the RG, i.e., the diffusion terms~$\propto p^2$ and the
rescaling of the fields including the anomalous dimensions. The next two lines
give one-loop contributions; if the remaining terms are dropped,
Eq.~\eqref{eq:(2.1)} reduces to the one-loop form of the Wetterich equation.
The remaining terms originate from the gauge-covariant diffusion of the
gradient flow and are peculiar to the GFERG. They are explicitly proportional
to~$e_\tau$ or~$e_\tau^2$, and contribute up to two and three loops,
respectively. Therefore, \emph{all the two- and three-loop contributions
in~Eq.~\eqref{eq:(2.1)} are proportional to the gauge coupling~$e_\tau$}; this
property holds independently of the ansatz and will be crucial
in~Sect.~\ref{sec:4}.

A remarkable feature of the GFERG in the $U(1)$ gauge theory with the gauge
fixing is that the RG flow according to~Eqs.~\eqref{eq:(2.1)}
and~\eqref{eq:(2.3)} preserves the following simple
structure~\cite{Sonoda:2022fmk},\footnote{%
Ref.~\cite{Sonoda:2025tyu} is a readable exposition of the GFERG.}
\begin{equation}
   \mathit{\Gamma}_\tau[\mathcal{A},\Bar{\Psi},\Psi]
   =\mathit{\Gamma}_\tau^{\text{inv}}[\mathcal{A},\Bar{\Psi},\Psi]
   -\frac{1}{2\xi_\tau}\int_k\,
   e^{k^2}\mathcal{A}_\mu(-k)e^{k^2}\mathcal{A}_\nu(k)k_\mu k_\nu,
\label{eq:(2.5)}
\end{equation}
where the part~$\mathit{\Gamma}_\tau^{\text{inv}}$ is invariant under the local
$U(1)$ gauge transformation,
\begin{align}
   \delta\mathcal{A}_\mu(k)&=ik_\mu\chi(k),&&
\notag\\
   \delta\Psi(p)&=ie_\tau\int_k\,\chi(k)\Psi(p-k),&
   \delta\Bar{\Psi}(-p)
   &=-ie_\tau\int_k\,\chi(k)\Bar{\Psi}(-p-k).
\label{eq:(2.6)}
\end{align}
This is a consequence of the gauge Ward--Takahashi (WT) identity that the
GFERG equation preserves. In particular, the nontransverse part of the photon
two-point function is completely fixed by~Eq.~\eqref{eq:(2.5)}, as in the
conventional 1PI effective action in (perturbative) QED. Since the gauge
transformation~\eqref{eq:(2.6)} takes the \emph{conventional\/} form, it is
straightforward to write down possible local terms
of~$\mathit{\Gamma}_\tau^{\text{inv}}$.

In this paper, we study the RG flow of the following ansatz, i.e., a
truncation of the 1PI Wilson action\footnote{%
In our convention, for a Euclidean action~$S$, the Boltzmann factor is~$e^S$,
instead of~$e^{-S}$.}
\begin{align}
   \mathit{\Gamma}_\tau[\mathcal{A},\Bar\Psi,\Psi]
   &=-\frac{1}{2}\int_k\,e^{k^2}\mathcal{A}_\mu(-k)e^{k^2}\mathcal{A}_\nu(k)
   \left[(k^2\delta_{\mu\nu}-k_\mu k_\nu)\mathcal{K}_\tau(k^2)
   +\frac{1}{\xi_\tau}k_\mu k_\nu\right]
\nonumber\\
   &\qquad{}
   -\int_p\,\Bar{\Psi}_{-1}(-p)(\Slash{p}+im_\tau)\Psi_{-1}(p)
   +e_\tau\int_{p,k}\,\Bar{\Psi}_{-1}(-p-k)e^{k^2}\Slash{\mathcal{A}}(k)\Psi_{-1}(p)
\nonumber\\
   &\qquad{}
   -2\kappa_\tau\int_{p,k}\,\Bar{\Psi}_{-1}(-p-k)
   \sigma_{\mu\nu}k_\mu e^{k^2}\mathcal{A}_\nu(k)\Psi_{-1}(p),
\label{eq:(2.7)}
\end{align}
where the first two lines are the QED ansatz
of~Refs.~\cite{Nagao:2026vaf,Nagao:2026gbq}; $\mathcal{K}_\tau(k^2)$ is the
kinetic function of the gauge field (we normalize $\mathcal{K}_\tau(0)=1$;
locality requires that $\mathcal{K}_\tau(k^2)$ be analytic and, in particular,
$k^2\mathcal{K}_\tau'(k^2)|_{k^2=0}=0$), $m_\tau$ is the mass parameter, and
$\kappa_\tau$ is the Pauli coupling constant. Our convention is
\begin{equation}
   \sigma_{\mu\nu}:=\frac{i}{2}[\gamma_\mu,\gamma_\nu],
\label{eq:(2.8)}
\end{equation}
which is hermitian. In coordinate space, the Pauli term reads
$i\kappa_\tau\Bar{\Psi}_{-1}\sigma_{\mu\nu}(e^{-\partial^2}\mathcal{F}_{\mu\nu})\Psi_{-1}$
with~$\mathcal{F}_{\mu\nu}=\partial_\mu\mathcal{A}_\nu-\partial_\nu\mathcal{A}_\mu$.
Together with the convention for the kinetic term, a real~$\kappa_\tau$
corresponds to a real Pauli coupling in Minkowski space. This choice matters:
with an anti-hermitian $\sigma_{\mu\nu}$, $\kappa_\tau\to i\kappa_\tau$ and the
sign of every $\kappa^2$ term below would be reversed. Like the mass term, the
Pauli term flips the chirality. Therefore, with the Pauli term, the mass
parameter receives an additive renormalization proportional
to~$e_\tau\kappa_\tau$ and~$\kappa_\tau^2$ even for~$m_\tau=0$. Moreover, the
fermion propagator of the GFERG, $h_F(p)=i/[e^{-2p^2}+i(\Slash{p}+im)]$
[Eq.~(3.17) of~Ref.~\cite{Nagao:2026vaf}], is not chirally symmetric: the term
$e^{-2p^2}$ acts like a Wilson term (cf.\ the Ginsparg--Wilson
relation~\cite{Ginsparg:1981bj} discussed
in~Refs.~\cite{Sonoda:2022fmk,Miyakawa:2021hcx}). Consequently, the pure gauge
interaction generates the Pauli term even for~$m_\tau=0$ (Sect.~\ref{sec:5}).
This is of the same origin as the additive renormalization of the mass
parameter: in the GFERG, chiral symmetry is realized only in a modified,
Ginsparg--Wilson-like form, which our ansatz does not fulfill. Since
$\kappa_\tau$ is irrelevant at the Gaussian fixed point, the induced value is
nonuniversal, like the critical mass, and it does not affect our conclusions
on the fixed-point structure; in particular, it plays no role in the subspace
$e_\tau=0$ studied in~Sect.~\ref{sec:4}.

The $-1$~variables $\Psi_{-1}$ and~$\Bar{\Psi}_{-1}$ in~Eq.~\eqref{eq:(2.7)} are
defined by the gradient flow equations (written in coordinate space)
\begin{align}
   \partial_t\mathcal{A}_{t\mu}(x)
   &=\partial_\nu
   \left[
   \partial_\nu\mathcal{A}_{t\mu}(x)-\partial_\mu\mathcal{A}_{t\nu}(x)
   \right]
   +\alpha_0\partial_\mu\partial_\nu\mathcal{A}_{t\nu}(x),&
   \mathcal{A}_{0\mu}(x)&=\mathcal{A}_\mu(x),
\notag\\
   \partial_t\Psi_t(x)
   &=\left\{
   \left[\partial_\mu-ie\mathcal{A}_{t\mu}(x)\right]^2
   +ie\alpha_0\left[\partial_\mu\mathcal{A}_{t\mu}(x)\right]
   \right\}\Psi_t(x),&
   \Psi_0(x)&=\Psi(x),
\notag\\
   \partial_t\Bar{\Psi}_t(x)
   &=\left\{
   \left[\partial_\mu+ie\mathcal{A}_{t\mu}(x)\right]^2
   -ie\alpha_0\left[\partial_\mu\mathcal{A}_{t\mu}(x)\right]
   \right\}\Bar{\Psi}_t(x),&
   \Bar{\Psi}_0(x)&=\Bar{\Psi}(x),
\label{eq:(2.9)}
\end{align}
by solving \emph{backward\/} from~$t=0$ to~$t=-1$. Equation~\eqref{eq:(2.9)}
is nothing but the gradient flow equations
in~Refs.~\cite{Luscher:2010iy,Luscher:2011bx,Luscher:2013cpa} specialized to
the $U(1)$ gauge theory. Here, $\alpha_0$ is a gauge parameter of the gradient
flow~\cite{Luscher:2010iy}: the term proportional to~$\alpha_0$ damps the
longitudinal mode of the flowed gauge field, which is not damped by the
gradient flow itself, and corresponds to an infinitesimal gauge transformation
of the flowed fields; gauge-invariant quantities do not depend on~$\alpha_0$.
In~Eq.~\eqref{eq:(2.9)}, the parameter~$e$ is regarded as a fixed constant and,
after solving the equations from~$t=0$ to~$t=-1$, $e$~is replaced by~$e_\tau$.

We use the $-1$~variables in the ansatz~\eqref{eq:(2.7)} because,
for~$e_\tau\to0$, $\mathcal{K}_\tau(k^2)\to1$, and~$\kappa_\tau\to0$
(and~$m_\tau=0$), Eq.~\eqref{eq:(2.7)} reduces to the Gaussian fixed point of
the GFERG~\cite{Sonoda:2022fmk}. We regard this inclusion of the Gaussian
fixed point as a minimal requirement that any ansatz for the 1PI Wilson action
should fulfill. Moreover, it can also be seen that, under the gauge
transformation~\eqref{eq:(2.6)}, defining the gauge transformation
function~$\chi_t(k)$ by
\begin{equation}
   \partial_t\chi_t(k)=-\alpha_0k^2\chi_t(k),\qquad
   \chi_0(k)=\chi(k),
\label{eq:(2.10)}
\end{equation}
the $-1$~variables transform as
\begin{align}
   \delta\mathcal{A}_{-1\mu}(k)&=ik_\mu\chi_{-1}(k),&&
\notag\\
   \delta\Psi_{-1}(p)&=ie_\tau\int_k\,\chi_{-1}(k)\Psi_{-1}(p-k),&
   \delta\Bar{\Psi}_{-1}(-p)
   &=-ie_\tau\int_k\,\chi_{-1}(k)\Bar{\Psi}_{-1}(-p-k).
\label{eq:(2.11)}
\end{align}
Therefore, the ansatz~\eqref{eq:(2.7)} is, apart from the gauge-fixing term
whose form is determined by~Eq.~\eqref{eq:(2.5)}, invariant under the gauge
transformation~\eqref{eq:(2.6)}. In particular, the Pauli term is gauge
invariant by itself, and Eq.~\eqref{eq:(2.3)} holds as without the Pauli term.

For~$\alpha_0=1$, which we set throughout this paper, $\mathcal{A}_{-1\mu}(x)$
is simply given by
\begin{equation}
   \mathcal{A}_{-1\mu}(x)=e^{-\partial^2}\mathcal{A}_\mu(x)
   \Rightarrow
   \mathcal{A}_{-1\mu}(k)=e^{k^2}\mathcal{A}_\mu(k),
\label{eq:(2.12)}
\end{equation}
whereas $\Psi_{-1}(x)$ and~$\Bar{\Psi}_{-1}(x)$ are of infinite order
in~$\mathcal{A}_\mu(x)$ through~Eq.~\eqref{eq:(2.9)}; we write
\begin{align}
   &\Psi_{-1}(p)
\notag\\
   &=\sum_{n\geq0}e_\tau^n\int_{k_1,\dotsc,k_n}\,
   \Hat{w}_n(p;k_1,\dotsc,k_n)_{\mu_1\dotsb\mu_n}
   \mathcal{A}_{\mu_1}(k_1)\dotsb\mathcal{A}_{\mu_n}(k_n)
   \Psi(p-k_1-\dotsb-k_n),
\notag\\
   &\Bar{\Psi}_{-1}(p)
\notag\\
   &=\sum_{n\geq0}(-e_\tau)^n\int_{k_1,\dotsc,k_n}\,
   \Hat{w}_n(p;k_1,\dotsc,k_n)_{\mu_1\dotsb\mu_n}
   \mathcal{A}_{\mu_1}(k_1)\dotsb\mathcal{A}_{\mu_n}(k_n)
   \Bar{\Psi}(p-k_1-\dotsb-k_n),
\label{eq:(2.13)}
\end{align}
where the functions~$\Hat{w}_n$ contain the exponential factors that result
from solving the free part of~Eq.~\eqref{eq:(2.9)},
$\partial_t\Psi_t=\partial^2\Psi_t$, backward to~$t=-1$. For instance,
for~$e_\tau=0$, $\Psi_{-1}(p)=e^{p^2}\Psi(p)$, i.e., $\Hat{w}_0(p)=e^{p^2}$, in the
same way as~$\mathcal{A}_{-1\mu}(k)=e^{k^2}\mathcal{A}_\mu(k)$
in~Eq.~\eqref{eq:(2.12)}. A general algorithm to obtain the
functions~$\Hat{w}_n(p;k_1,\dotsc,k_n)_{\mu_1\dotsb\mu_n}$ is given in~Appendix~B
of~Ref.~\cite{Nagao:2026gbq}.

By using Eq.~\eqref{eq:(2.13)} in~Eq.~\eqref{eq:(2.7)}, we define vertex
functions~$\Hat{V}_{\mu_1\dotsb\mu_n}(-p-k_1-\dotsb-k_n,p;k_1,\dotsc,k_n)$
and~$\Hat{P}_{\mu_1\dotsb\mu_n}(-p-k_1-\dotsb-k_n,p;k_1,\dotsc,k_n)$ by
\begin{align}
   &-\int_p\,
   \Bar{\Psi}_{-1}(-p)(\Slash{p}+im_\tau)\Psi_{-1}(p)
   +e_\tau\int_{p,k}\,
   \Bar{\Psi}_{-1}(-p-k)e^{k^2}\Slash{\mathcal{A}}(k)\Psi_{-1}(p)
\notag\\
   &=-\int_p\,\Bar{\Psi}(-p)\Hat{V}(p)\Psi(p)
\notag\\
   &\qquad{}
   +\sum_{n=1}^\infty e_\tau^n\int_{p,k_1,\dotsc,k_n}
   \Bar{\Psi}(-p-k_1-\dotsb-k_n)
\notag\\
   &\qquad\qquad{}
   \times
   \Hat{V}_{\mu_1\dotsb\mu_n}(-p-k_1-\dotsb-k_n,p;k_1,\dotsc,k_n)
   \mathcal{A}_{\mu_1}(k_1)\dotsb\mathcal{A}_{\mu_n}(k_n)
   \Psi(p),
\label{eq:(2.14)}
\end{align}
and
\begin{align}
   &-2\kappa_\tau\int_{p,k}\,
   \Bar{\Psi}_{-1}(-p-k)\sigma_{\mu\nu}k_\mu e^{k^2}\mathcal{A}_\nu(k)\Psi_{-1}(p)
\notag\\
   &=\frac{\kappa_\tau}{e_\tau}
   \sum_{n=1}^\infty e_\tau^n\int_{p,k_1,\dotsc,k_n}
   \Bar{\Psi}(-p-k_1-\dotsb-k_n)
\notag\\
   &\qquad\qquad{}
   \times
   \Hat{P}_{\mu_1\dotsb\mu_n}(-p-k_1-\dotsb-k_n,p;k_1,\dotsc,k_n)
   \mathcal{A}_{\mu_1}(k_1)\dotsb\mathcal{A}_{\mu_n}(k_n)
   \Psi(p).
\label{eq:(2.15)}
\end{align}
The vertex functions~$\Hat{V}_{\mu_1\dotsb\mu_n}$ arise from the ansatz without the
Pauli term (also used in~Refs.~\cite{Nagao:2026vaf,Nagao:2026gbq}) and
$\Hat{P}_{\mu_1\dotsb\mu_n}$ from the Pauli term.

Now, substituting Eq.~\eqref{eq:(2.7)} into the GFERG
equation~\eqref{eq:(2.1)} and expanding with respect to~$\mathcal{A}$, we have
\begin{align}
   &-\frac{1}{2}\int_k\,e^{k^2}\mathcal{A}_\mu(-k)e^{k^2}\mathcal{A}_\nu(k)
   \left[(k^2\delta_{\mu\nu}-k_\mu k_\nu)\partial_\tau\mathcal{K}_\tau(k^2)
   -2(\epsilon+\gamma_A)\frac{1}{\xi_\tau}k_\mu k_\nu\right]
\nonumber\\
   &\qquad{}
   -\int_p\,e^{p^2}\Bar{\Psi}(-p)i\frac{d}{d\tau}m_\tau e^{p^2}\Psi(p)
\nonumber\\
   &\qquad{}
   -\gamma_Ae_\tau
   \int_{p,k}\,e^{(p+k)^2}\Bar{\Psi}(-p-k)e^{k^2}
   \Slash{\mathcal{A}}(k)e^{p^2}\Psi(p)
\nonumber\\
   &\qquad{}
   -2\frac{d}{d\tau}\kappa_\tau\int_{p,k}\,e^{(p+k)^2}\Bar{\Psi}(-p-k)
   \sigma_{\mu\nu}k_\mu e^{k^2}\mathcal{A}_\nu(k)e^{p^2}\Psi(p)
\notag\\
   &\qquad{}
   +\text{higher orders in~$\mathcal{A}$}
\notag\\
   &=-\frac{1}{2}\int_k\,e^{k^2}\mathcal{A}_\mu(-k)
   \biggl\{
   (k^2\delta_{\mu\nu}-k_\mu k_\nu)
   \left[
   -2(\epsilon+\gamma_A)\mathcal{K}_\tau(k^2)
   -2k^2\mathcal{K}_\tau'(k^2)
   \right]
\notag\\
   &\qquad\qquad\qquad\qquad\qquad\qquad\qquad{}
   +\mathcal{R}_{\mu\nu}^{\text{loop}}(k)
   -2(\epsilon+\gamma_A)\frac{1}{\xi_\tau}k_\mu k_\nu
   \biggr\}
   e^{k^2}\mathcal{A}_\nu(k)
\notag\\
   &\qquad{}
   -\int_p\,
   e^{p^2}\Bar{\Psi}(-p)
   \left[
   -2\gamma_\psi(\Slash{p}+im_\tau)+im_\tau
   +\mathcal{R}(p)
   \right]
   e^{p^2}\Psi(p)
\notag\\
   &\qquad{}
   +\int_{p,k}\,
   e^{(p+k)^2}\Bar{\Psi}(-p-k)
   \bigl[
   -(2\gamma_\psi+\gamma_A)e_\tau\gamma_\nu
   +2(1+2\gamma_\psi+\gamma_A)\kappa_\tau\sigma_{\mu\nu}k_\mu
\notag\\
   &\qquad\qquad\qquad\qquad\qquad\qquad\qquad{}
   +X_\nu(-k-p,p;k)\bigr]
   e^{k^2}\mathcal{A}_\nu(k)
   e^{p^2}\Psi(p)
\notag\\
   &\qquad{}
   +\text{higher orders in~$\mathcal{A}$},
\label{eq:(2.16)}
\end{align}
where we have used~Eq.~\eqref{eq:(2.3)}. In the terms linear in~$\mathcal{A}$,
we have shown only the Dirac structures that are relevant for the
projection~\eqref{eq:(2.21)} below. The fermion kinetic term
in~Eq.~\eqref{eq:(2.7)} also contributes to the terms linear in~$\mathcal{A}$
through~$\Hat{w}_1$ in~Eq.~\eqref{eq:(2.13)}; together with the gauge vertex,
it forms $e_\tau\Hat{V}_\nu(-p-k,p;k)$ in~Eq.~\eqref{eq:(2.14)}, and its
$\tau$-derivative produces terms proportional to~$de_\tau/d\tau$
and~$dm_\tau/d\tau$ on the left-hand side. At~$p=0$, however, all these
additional terms are proportional to~$k_\nu$ [cf.\ the vertex
$\gamma_\nu+2im_\tau k_\nu F(-2k^2)$ ($F(x)=(e^x-1)/x$) of the $-1$ variables;
Ref.~\cite{Nagao:2026gbq}], and therefore vanish upon the contraction
with~$f_\nu$ in~Eq.~\eqref{eq:(2.21)}.

Since the structure~\eqref{eq:(2.5)} is preserved by the flow, the loop
part~$\mathcal{R}^{\text{loop}}_{\mu\nu}(k)$ is transverse. Therefore, we set the
loop contributions as
\begin{align}
   \mathcal{R}_{\mu\nu}^{\text{loop}}(k)
   &=(k^2\delta_{\mu\nu}-k_\mu k_\nu)\mathcal{L}(k^2),
\notag\\
   \mathcal{R}(p)&=a\Slash{p}+ib+O(p^2),
\notag\\
   X_\mu(-k,0;k)
   &=c_\gamma\gamma_\mu+c_\sigma\sigma_{\mu\nu}k_\nu+\text{other Dirac structures}.
\label{eq:(2.17)}
\end{align}
Then, using the locality~$k^2\mathcal{K}_\tau'(k^2)|_{k^2=0}=0$,
Eq.~\eqref{eq:(2.16)} yields
\begin{align}
   \partial_\tau\mathcal{K}_\tau(k^2)
   &=-2[(\epsilon+\gamma_A)\mathcal{K}_\tau(k^2)+k^2\mathcal{K}_\tau'(k^2)]
   +\mathcal{L}(k^2),
\notag\\
   \frac{d}{d\tau}m_\tau
   &=(1-a)m_\tau+b,
\notag\\
   \frac{d}{d\tau}\kappa_\tau
   &=-(1+\gamma_A+2\gamma_\psi)\kappa_\tau+\frac{1}{2}c_\sigma,
\label{eq:(2.18)}
\end{align}
and
\begin{align}
   \gamma_A&=-\epsilon+\frac{1}{2}\mathcal{L}(k^2=0),
\notag\\
   \gamma_\psi&=\frac{a}{2},
\notag\\
   c_\gamma&=2\gamma_\psi e_\tau=ae_\tau.
\label{eq:(2.19)}
\end{align}

When there is no Pauli term, the explicit forms of~$\mathcal{L}(k^2)$
and~$\mathcal{R}(p)$ in~Eq.~\eqref{eq:(2.17)} in terms of the vertex
functions~$\Hat{V}_{\mu_1\dotsb\mu_n}$ are given in~Appendix~A
of~Ref.~\cite{Nagao:2026gbq}. These expressions follow from the GFERG
equation~\eqref{eq:(2.1)} for any ansatz whose fermion part is expanded as
in~Eq.~\eqref{eq:(2.14)}. The effect of the Pauli term can therefore be
incorporated by simply replacing
\begin{equation}
   \Hat{V}_{\mu_1\dotsb\mu_n}
   \to\Hat{V}_{\mu_1\dotsb\mu_n}
   +\frac{\kappa_\tau}{e_\tau}\Hat{P}_{\mu_1\dotsb\mu_n}
\label{eq:(2.20)}
\end{equation}
in those expressions, including the vertex functions in the combinations such
as~$\Hat{\mathcal{A}}_{\mu\nu}$, $\Hat{\mathcal{B}}_{\mu\nu\rho}$,
and~$\Hat{\mathcal{C}}_{\mu\nu\rho\sigma}$ there. The anomalous dimensions
$\gamma_A$~and $\gamma_\psi$ enter $\mathcal{L}(k^2)$ and~$\mathcal{R}(p)$
linearly and are determined algebraically as in~Eq.~(3.3)
of~Ref.~\cite{Nagao:2026gbq}.

$X_\nu$ in~Eq.~\eqref{eq:(2.16)} is the loop part of the fermion--photon
three-point function; it contains contributions from the gauge interaction
alone as well as those from the Pauli term. The coefficients $c_\gamma$
and~$c_\sigma$ in~Eq.~\eqref{eq:(2.17)} can be obtained from~$X_\mu(-k,0;k)$ by
\begin{equation}
   c_\gamma=\frac{1}{\tr(1)}\tr(f_\alpha\gamma_\alpha f_\mu X_\mu),\qquad
   c_\sigma=\lim_{k\to0}
   \frac{1}{\tr(1)k^2}\tr(f_\alpha k_\beta\sigma_{\alpha\beta}f_\mu X_\mu),
\label{eq:(2.21)}
\end{equation}
where $f_\mu$ is a unit vector satisfying $f\cdot k=0$ and~$f^2=1$, and
$\tr(1)$ is the trace of the unit matrix in Dirac space. Without the Pauli
term, the three-point function~$X_\mu$ need not be computed, since the flow
of~$e_\tau$ is fixed by the WT identity [Eq.~\eqref{eq:(2.3)}] and the
projection onto the structure~$\gamma_\mu$ merely reproduces it
through~$c_\gamma=ae_\tau$. The Pauli coupling~$\kappa_\tau$, on the other hand,
is not constrained by the WT identity, and its flow requires~$c_\sigma$. No
closed expression analogous to those of~$\mathcal{L}(k^2)$ and~$\mathcal{R}(p)$
in~Ref.~\cite{Nagao:2026gbq} is available for~$X_\nu$, and we compute it
numerically as sketched in~Sect.~\ref{sec:5.1}. We use the last relation
in~Eq.~\eqref{eq:(2.19)}, $c_\gamma=ae_\tau$, as a consistency check of the
numerical computation.

\section{Leading large-$N_f$ approximation}
\label{sec:3}
\begin{figure}[htbp]
\centering
\includegraphics[width=0.9\textwidth]{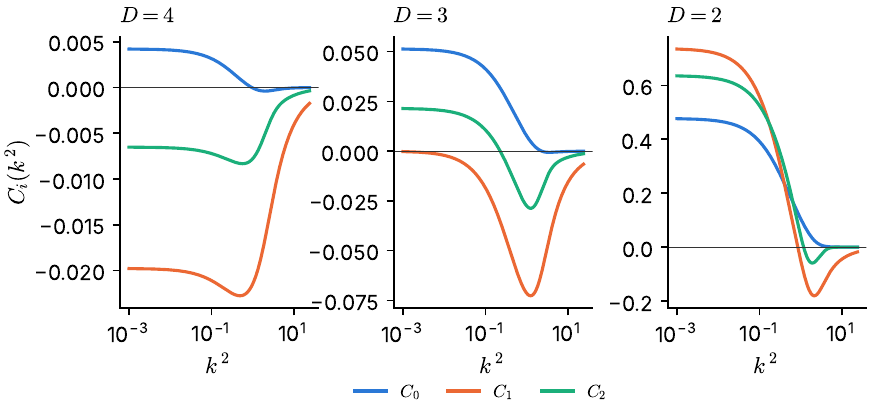}
\caption{The functions~$C_i(k^2)$ in the leading large-$N_f$ approximation
for~$m=0$.}
\label{fig:1}
\end{figure}

In this section, we consider $N_f$ flavors of the fermion and study the flow
in the leading order of the $1/N_f$ expansion. We define
\begin{equation}
   \Tilde{e}^2:=e_\tau^2\tr(1)N_f,\qquad
   \Tilde{\kappa}:=\kappa_\tau\sqrt{\tr(1)N_f},
\label{eq:(3.1)}
\end{equation}
and take the limit~$N_f\to\infty$ with $\Tilde{e}$ and~$\Tilde{\kappa}$
fixed~\cite{PalanquesMestre:1983zy,Gracey:1996he,Holdom:2010qs}; for the case
without the Pauli term, see also Sect.~4.2 of~Ref.~\cite{Nagao:2026vaf}. Since
a closed fermion loop gives a factor~$N_f$, the one-loop photon two-point
function is of~$O(1)$ in terms of~$\Tilde{e}$ and~$\Tilde{\kappa}$, whereas
the multi-loop terms of the photon two-point function and all the loop terms
of the fermion two- and three-point functions are suppressed by~$1/N_f$.
Therefore, $\mathcal{L}(k^2)$ in~Eq.~\eqref{eq:(2.17)} reduces to its one-loop
part, and $a$, $b$, and~$c_\sigma$ drop out. Then, Eqs.~\eqref{eq:(2.3)},
\eqref{eq:(2.18)}, and~\eqref{eq:(2.19)} become
\begin{align}
   \frac{d}{d\tau}\Tilde{e}&=-\gamma_A\Tilde{e},
\notag\\
   \frac{d}{d\tau}\Tilde{\kappa}&=-(1+\gamma_A)\Tilde{\kappa},
\notag\\
   \frac{d}{d\tau}m_\tau&=m_\tau,
\notag\\
   \partial_\tau\mathcal{K}_\tau(x)
   &=-2\left[(\epsilon+\gamma_A)\mathcal{K}_\tau(x)+x\mathcal{K}_\tau'(x)\right]
   +\mathcal{C}(x),
\label{eq:(3.2)}
\end{align}
where $x:=k^2$ and
\begin{align}
   \gamma_A&=-\epsilon+\frac{1}{2}\mathcal{C}(0),
\notag\\
   \mathcal{C}(x)&=\Tilde{e}^2C_0(x)+\Tilde{e}\Tilde{\kappa}C_1(x)
   +\Tilde{\kappa}^2C_2(x),
\notag\\
   \gamma_\psi&=0.
\label{eq:(3.3)}
\end{align}
The functions~$C_i(x)$ are the one-loop part of~$\mathcal{L}(x)$ given
in~Appendix~A of~Ref.~\cite{Nagao:2026gbq} with the
replacement~\eqref{eq:(2.20)}, divided by~$\tr(1)N_f$. Since the one-loop
photon two-point function contains no photon propagator, $C_i(x)$ do not depend
on~$\mathcal{K}_\tau$; they depend only on~$m_\tau$. Hence, the couplings
$\Tilde{e}$, $\Tilde{\kappa}$, and~$m_\tau$ obey a closed system of ordinary
differential equations, and $\mathcal{K}_\tau$ is driven by them without
feedback. In particular, Eq.~\eqref{eq:(3.2)}
gives~$\Tilde{\kappa}/\Tilde{e}\propto e^{-\tau}$. Since $m_\tau$ is relevant and
vanishes at all the fixed points below, we show~$C_i$ for~$m=0$
in~Table~\ref{table:1} and~Fig.~\ref{fig:1}.
\begin{table}[htbp]
\centering\small
\begin{tabular}{lccc}\toprule
& $C_0(0)$ & $C_1(0)$ & $C_2(0)$\\
\midrule
$D=4$ & $0.0042215\,(=1/24\pi^2)$ & $-0.019769$ & $-0.0065052$\\
$D=3$ & $0.051421$ & $|C_1(0)|<10^{-5}$ & $0.021526$\\
$D=2$ & $0.47745\,(=3/2\pi)$ & $0.73692$ & $0.63660\,(=2/\pi)$\\
\bottomrule
\end{tabular}
\caption{The coefficients~$C_i(0)$ of the photon two-point function in the
leading large-$N_f$ approximation for~$m=0$. $C_0(0)$ reproduces the values
of~Ref.~\cite{Nagao:2026vaf}.}
\label{table:1}
\end{table}

From the first two relations in~Eq.~\eqref{eq:(3.2)}, a fixed point
with~$\Tilde{e}\neq0$ requires~$\gamma_A=0$, while one
with~$\Tilde{\kappa}\neq0$ requires~$\gamma_A=-1$. Therefore, $\Tilde{e}$
and~$\Tilde{\kappa}$ cannot be nonzero simultaneously at a fixed point. The
fixed points and the eigenvalues of the linearized flow are listed
in~Table~\ref{table:2}. For~$\mathcal{K}_\tau$, the linearization of the last
relation in~Eq.~\eqref{eq:(3.2)} shows that the mode~$x^n$ has the
eigenvalue~$-2(\epsilon+\gamma_A+n)$; the mode with~$n=0$ is excluded, since
$\mathcal{K}_\tau(0)=1$ is fixed by the normalization of~$\mathcal{A}$.
\begin{table}[htbp]
\centering\small
\begin{tabular}{llcccl}
\toprule
fixed point & coordinates & $\lambda_\kappa$ & $\lambda_e$ & $\lambda_\xi$
& $\mathcal{K}$ modes~$x^n$ ($n\geq1$)\\
\midrule
Gaussian & $\Tilde{e}=\Tilde{\kappa}=0$ & $\epsilon-1$ & $\epsilon$ & $0$
& $-2n$\\
IR ($D<4$) & $\Tilde{e}_*^2=2\epsilon/C_0(0)$, $\Tilde{\kappa}=0$ & $-1$
& $-2\epsilon$ & $2\epsilon$ & $-2(n+\epsilon)$\\
Pauli ($D=4$) & $\Tilde{e}=0$, $\Tilde{\kappa}_*^2=-2/C_2(0)$ & $2$ & $1$
& $-2$ & $-2(n-1)$\\
\bottomrule
\end{tabular}
\caption{Fixed points in the leading large-$N_f$ approximation and the
eigenvalues of the linearized flow; a positive eigenvalue corresponds to a
relevant direction, and $\lambda_m=1$ at all the fixed points. The IR fixed
point has~$\Tilde{e}_*^2=19.45$ ($D=3$) and~$4\pi/3$ ($D=2$), in agreement
with~Ref.~\cite{Nagao:2026gbq}.}
\label{table:2}
\end{table}

\begin{figure}[htbp]
\centering
\includegraphics[width=0.75\textwidth]{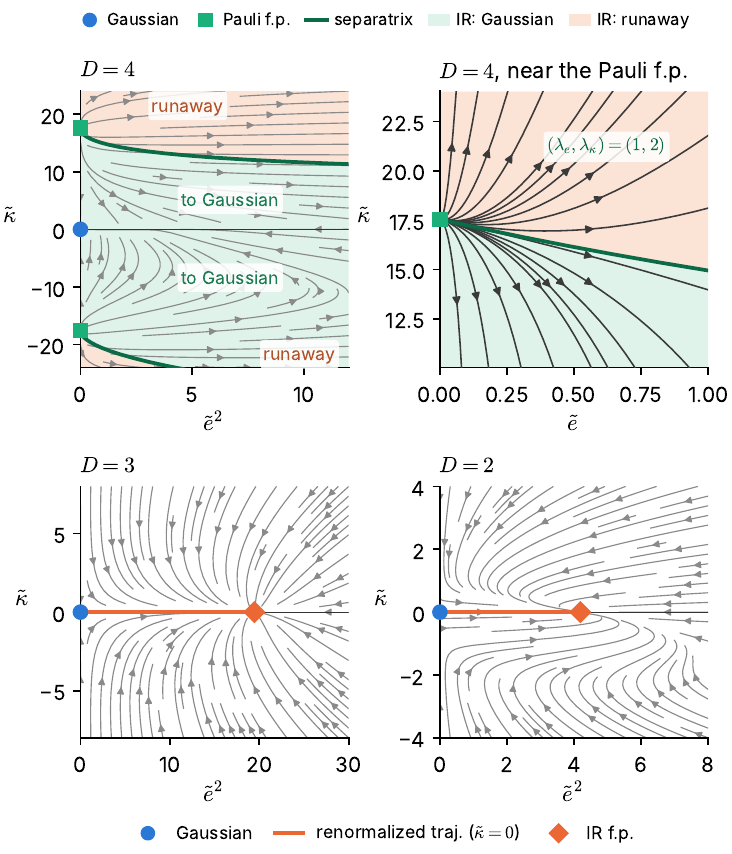}
\caption{RG flow in the leading large-$N_f$ approximation; the arrows point
toward the IR. Top left: $D=4$ in the~$(\Tilde{e}^2,\Tilde{\kappa})$ plane;
the shading indicates whether a trajectory approaches the Gaussian fixed point
(green) or runs away (orange) in the IR, and the dark green curves are the
separatrices. Top right: the neighborhood of the Pauli fixed point
with~$\Tilde{\kappa}_*>0$ in the~$(\Tilde{e},\Tilde{\kappa})$ plane; all the
trajectories emanate from the fixed point, tangentially to the eigendirection
with~$\lambda_e=1$. Bottom: $D=3$ (left) and~$D=2$ (right) in
the~$(\Tilde{e}^2,\Tilde{\kappa})$ plane; the orange line is the renormalized
trajectory from the Gaussian to the IR fixed point.}
\label{fig:2}
\end{figure}

In~$D=4$, since $C_2(0)<0$, there exists a fixed point with a nonzero Pauli
coupling, which we call the \emph{Pauli fixed point},
\begin{equation}
   \Tilde{e}_*=0,\qquad
   \Tilde{\kappa}_*^2=-\frac{2}{C_2(0)}=307.4,\qquad
   \gamma_A=-1.
\label{eq:(3.4)}
\end{equation}
Linearizing the first two relations in~Eq.~\eqref{eq:(3.2)}
around~Eq.~\eqref{eq:(3.4)} with~$\gamma_A$ in~Eq.~\eqref{eq:(3.3)}, we find
that the stability matrix in~$(\delta\Tilde{e},\delta\Tilde{\kappa})$ is
triangular with the eigenvalues~$1$ and~$2$, where we have
used~$\Tilde{\kappa}_*^2C_2(0)=-2$. Together with the mass ($\lambda_m=1$), the
Pauli fixed point thus has three relevant directions. In particular, the gauge
coupling is relevant, i.e., asymptotically free, at this fixed point, in
accordance with~$\lambda_e=-\gamma_A^*=1$ [Eq.~\eqref{eq:(2.4)}]. The fixed
point~$\mathcal{C}$ of~Ref.~\cite{Gies:2020xuh} also has three relevant
directions, including the gauge coupling. The mechanism is, however, different.
There, the photon anomalous dimension vanishes and the fermion anomalous
dimension is~$\gamma_\psi=-1/2$,\footnote{%
The anomalous dimensions $\eta_A$ and~$\eta_\psi$ of~Ref.~\cite{Gies:2020xuh}
correspond to our~$2\gamma_A$ and~$2\gamma_\psi$, respectively, as seen by
comparing their flow of the Pauli coupling,
$\partial_t\kappa=(1+\eta_A/2+\eta_\psi)\kappa+\dotsb$, where~$t=-\tau$,
with~Eq.~\eqref{eq:(2.18)}.}
which cancels the canonical dimension of the Pauli coupling. In our case,
$\gamma_\psi=0$, and the relevance of both the Pauli and the gauge couplings is
due to the vacuum polarization, $\gamma_A^*=-1$. Since both~$\Tilde{e}$
and~$\Tilde{\kappa}$ are relevant at the Pauli fixed point, every trajectory in
its neighborhood in the~$(\Tilde{e},\Tilde{\kappa})$ plane emanates from the
fixed point, i.e., approaches it toward the UV; such a trajectory is a
renormalized trajectory, defining a continuum limit with the Pauli fixed point
as its UV limit. We have checked numerically that this holds not only in the
neighborhood of the fixed point: in the region shown in~Fig.~\ref{fig:2},
every trajectory with~$\Tilde{\kappa}\neq0$ approaches one of the Pauli fixed
points toward the UV. Toward the IR, these trajectories are divided by a
separatrix into two classes: those on the side of~$\Tilde{\kappa}=0$ approach
the Gaussian fixed point, along which $\Tilde{e}$ vanishes only
logarithmically, whereas the others run away to strong coupling
(Fig.~\ref{fig:2}). Along the former, a continuum limit in which the gauge
coupling is nonzero at any finite scale would be obtained, apart from the
resonance of~$\mathcal{K}_\tau$ discussed below.

Two remarks are in order. First, at~$\gamma_A=-1$, the eigenvalue of the
mode~$x^1$ of~$\mathcal{K}_\tau$, i.e., of the
operator~$\mathcal{F}_{\mu\nu}\partial^2\mathcal{F}_{\mu\nu}$, vanishes
(Table~\ref{table:2}), whereas the inhomogeneous term does not:
\begin{equation}
   \partial_\tau\mathcal{K}_\tau'(0)
   =\Tilde{\kappa}_*^2C_2'(0)\simeq-2.56.
\label{eq:(3.5)}
\end{equation}
Hence, $\mathcal{K}_\tau'(0)$ keeps flowing linearly in~$\tau$, and the Pauli
fixed point is a fixed point of the couplings only, not of the full system
including the function~$\mathcal{K}_\tau$. A stationary solution would contain
the term~$\frac{1}{2}\Tilde{\kappa}_*^2C_2'(0)x\ln x$, which is incompatible
with the locality. This resonance might be resolved at finite~$N_f$, if
$\gamma_A$ at the fixed point deviates from~$-1$; see, however,
Sect.~\ref{sec:4}. Second, in the functional RG analysis
of~Refs.~\cite{Gies:2020xuh,Gies:2024ugz}, the fixed points arise already
for~$N_f=1$, and the fermion fluctuations with a large fermion anomalous
dimension should play an important role there. Such effects are suppressed in
the leading large-$N_f$ approximation; we examine them in~Sect.~\ref{sec:4}.

In~$D=3$, since $C_2(0)>0$, there is no Pauli fixed point, and $\kappa_\tau$ is
irrelevant at both the Gaussian and the IR fixed points. In~$D=2$,
$\kappa_\tau$ is marginal at the Gaussian fixed point; it is marginally
irrelevant, since
$d\Tilde{\kappa}/d\tau=-\Tilde{\kappa}^3/\pi$ for~$\Tilde{e}=0$.\footnote{%
This is understood as follows: for~$D=2$ and~$\Tilde{e}=0$, $\mathcal{F}_{12}$
is an auxiliary field, and integrating it out yields a four-fermion
interaction in the pseudoscalar channel,
$-2\kappa_\tau^2(\Bar{\psi}\gamma_5\psi)^2$, with the repulsive sign. The above
beta function coincides with that of this four-fermion interaction in the
leading large-$N$ approximation~\cite{Gross:1974jv}, with the sign opposite to
that of the asymptotically free Gross--Neveu model.}

\section{Finite $N_f$: the subspace $e_\tau=0$}
\label{sec:4}
In this section, we go beyond the leading large-$N_f$ approximation in the
subspace~$e_\tau=0$, which describes fermions coupled to the photon only
through the Pauli term; the fixed points
of~Refs.~\cite{Gies:2020xuh,Gies:2022aiz,Gies:2024ugz} are located in this
subspace. As remarked below~Eq.~\eqref{eq:(2.3)}, all the two- and three-loop
contributions to the GFERG equation are proportional to~$e_\tau$. Since the
subspace~$e_\tau=0$ is preserved by~Eq.~\eqref{eq:(2.3)}, the right-hand side
of the GFERG equation~\eqref{eq:(2.16)} in this subspace is \emph{exactly of
one-loop order} for any~$N_f$. At one loop, the photon two-point function
contains a closed fermion loop and is proportional to~$N_f$, whereas the
fermion two- and three-point functions are independent of~$N_f$.

\subsection{Weak Pauli coupling}
\label{sec:4.1}
Table~\ref{table:3} classifies the one-loop terms at the Gaussian point
($m=0$, $\mathcal{K}=1$) by their $N_f$ dependence. The leading large-$N_f$
approximation in~Sect.~\ref{sec:3} keeps only the first line.
\begin{table}[htbp]
\centering\small
\begin{tabular}{llll}
\toprule
quantity & one-loop term & origin & $N_f$ dependence\\
\midrule
$\gamma_A$ & $\frac{1}{2}\tr(1)N_fC_2(0)\kappa^2\simeq-0.013N_f\kappa^2$
& vacuum polarization & $\propto N_f$\\
$2\gamma_\psi=a$ & $\simeq+0.061\kappa^2$ & fermion self-energy & $N_f^0$\\
$c_\sigma/(2\kappa)$ & $\simeq+0.003\kappa^2$ & vertex correction & $N_f^0$\\
$b$ & $\simeq-0.041\kappa^2$ & fermion self-energy (mass part) & $N_f^0$\\
\bottomrule
\end{tabular}
\caption{One-loop terms in the subspace~$e_\tau=0$ ($D=4$, $m=0$,
$\mathcal{K}=1$, and~$\tr(1)=4$).}
\label{table:3}
\end{table}
At the large-$N_f$ Pauli fixed point~\eqref{eq:(3.4)},
$\kappa_*^2=\Tilde{\kappa}_*^2/(\tr(1)N_f)\simeq77/N_f$, so the neglected terms
are of the size
\begin{equation}
   2\gamma_\psi\simeq\frac{4.7}{N_f},\qquad
   m_c\simeq0.04\,\kappa^2\simeq\frac{3}{N_f},
\label{eq:(4.1)}
\end{equation}
where $m_c=-b/(1-a)$ is the critical value of the mass parameter
[cf.~Eq.~\eqref{eq:(2.18)}]. These terms affect the flow of~$\kappa_\tau$ in
two ways. First, $2\gamma_\psi$ enters the
eigenvalue~$-(1+\gamma_A+2\gamma_\psi)$ in~Eq.~\eqref{eq:(2.18)} with the sign
opposite to~$\gamma_A$; for~$N_f=1$, $2\gamma_\psi$ alone
overcompensates~$\gamma_A=-1$. Second, since the Pauli term, together with the
Wilson-term-like part of~$h_F$, breaks the chirality (Sect.~\ref{sec:2}), the
critical surface moves to~$m_c\propto\kappa^2$. The fermion on the critical
surface then becomes heavy in units of the cutoff, and the vacuum polarization
that makes~$\gamma_A$ negative is suppressed (decoupling).

For small~$\kappa_\tau$, Table~\ref{table:3} gives
\begin{equation}
   \beta_\kappa:=\frac{d\kappa_\tau}{d\tau}
   =\kappa_\tau\left[-1+(0.0130N_f-0.0581)\kappa_\tau^2+O(\kappa_\tau^3)\right],
\label{eq:(4.2)}
\end{equation}
and the coefficient of~$\kappa_\tau^2$ is negative for~$N_f\leq4$. This
coefficient does not depend on the truncation: In the subspace~$e_\tau=0$, the
loop coefficients~$a$, $b$, $c_\sigma$, and~$\mathcal{L}$
in~Eq.~\eqref{eq:(2.17)} consist of one-loop integrals only, and $\gamma_A$,
$\gamma_\psi$, and~$c_\sigma/\kappa$ at~$O(\kappa^2)$ are given by the Pauli
vertex and the free propagators alone. Operators outside the ansatz, such as
the higher-derivative terms of the fermion two- and three-point functions and
the four-fermion operators, as well as the deviations~$m_c\propto\kappa^2$
and~$\mathcal{K}_\tau-1\propto\kappa^2$, are themselves generated
at~$O(\kappa^2)$ or higher, and affect~$\beta_\kappa/\kappa_\tau$ only at higher
orders in~$\kappa_\tau$. Therefore, for~$N_f\leq4$, the Pauli coupling becomes
more irrelevant as it grows from zero.

\begin{figure}[htbp]
\centering
\includegraphics[width=0.8\textwidth]{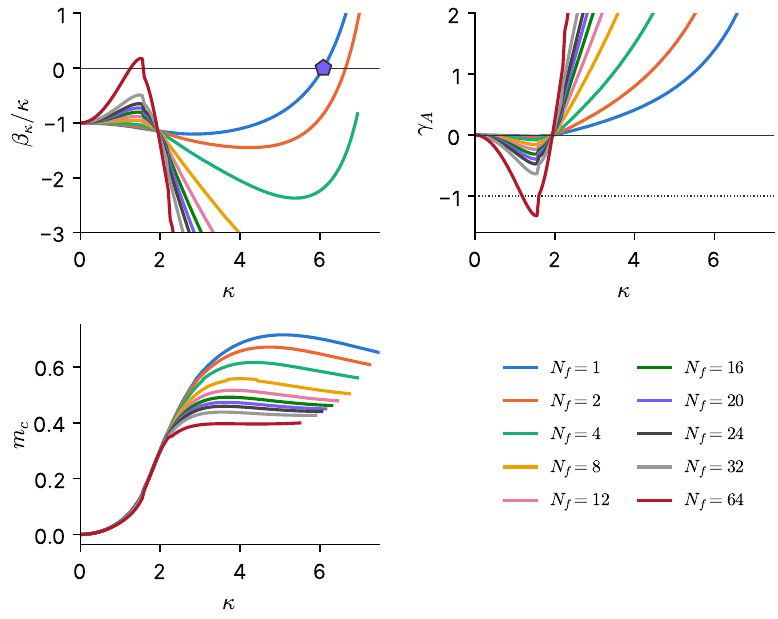}
\caption{$\beta_\kappa/\kappa$, $\gamma_A$, and~$m_c$ as functions of~$\kappa$
in the subspace~$e_\tau=0$ (exactly of one-loop order) for various~$N_f$ ($D=4$,
$\mathcal{K}=1$). The mass parameter is set to the value~$m_c(\kappa)$ at which
$dm_\tau/d\tau=0$, on the branch connected to~$m=0$. The zeros
of~$\beta_\kappa/\kappa$ are the fixed points; elsewhere, the curves do not
represent the flow on the critical surface (see text). For~$N_f=1$,
$\beta_\kappa/\kappa$ has a zero at~$\kappa\simeq6.1$ (pentagon in the
upper-left panel), which corresponds to the fixed point~\eqref{eq:(4.3)} found
in~Sect.~\ref{sec:4.2} (a similar zero appears at~$\kappa\simeq6.7$
for~$N_f=2$). Each curve ends where the branch can no longer be followed.}
\label{fig:3}
\end{figure}

Figure~\ref{fig:3} shows~$\beta_\kappa/\kappa$, $\gamma_A$, and~$m_c$ as
functions of~$\kappa$ for various~$N_f$. Here we set~$\mathcal{K}=1$ and tune
the mass parameter to its critical value~$m_c(\kappa)$, i.e., $dm_\tau/d\tau=0$,
on the branch connected to~$m=0$; a fixed point then corresponds to a zero
of~$\beta_\kappa/\kappa$. This reduces the search for fixed points to a
one-variable problem and allows us to compare various~$N_f$ in one figure.
Note that the curve~$dm_\tau/d\tau=0$ is not the critical surface in the strict
sense, i.e., the invariant curve on which $m_\tau$ does not run away; the two
coincide only at fixed points. Near the Gaussian fixed point, for instance,
where $d\kappa_\tau/d\tau\simeq-\kappa_\tau$ and
$dm_\tau/d\tau\simeq m_\tau-c\kappa_\tau^2$ with a constant~$c$, the former is
$m=c\kappa^2$. The latter is obtained by solving these equations:
with~$\kappa_\tau=\kappa_0e^{-\tau}$, the general solution
is~$m_\tau=c\kappa_\tau^2/3+Ce^{\tau}$, and $m_\tau$ does not run away only
for~$C=0$, i.e., on~$m=c\kappa^2/3$. Hence, the zeros of~$\beta_\kappa/\kappa$
in~Fig.~\ref{fig:3} are exactly the fixed points, but its values elsewhere do
not represent the flow on the critical surface. As $\kappa$ increases from
zero, two effects compete
in~$\beta_\kappa/\kappa=-(1+\gamma_A+2\gamma_\psi)+c_\sigma/(2\kappa)$. The vacuum
polarization makes $\gamma_A$ negative in proportion to~$N_f$ and
raises~$\beta_\kappa/\kappa$, whereas the $N_f$-independent terms,
mainly~$2\gamma_\psi$, lower it. In addition, the critical mass~$m_c(\kappa)$
grows with~$\kappa$, and the heavier fermion suppresses the vacuum
polarization; $\gamma_A$ thus returns to zero at~$\kappa\simeq1.95$ and becomes
positive beyond it. Since the vacuum polarization is proportional to~$N_f$ and
$m_c(\kappa)$ is almost independent of~$N_f$ for~$\kappa\lesssim2$, this point
is practically common to all~$N_f$. Consequently, a zero
of~$\beta_\kappa/\kappa$ caused by the vacuum polarization can appear only in
the window~$\kappa\lesssim1.9$, where $\beta_\kappa/\kappa$ has a maximum. This
maximum grows almost linearly in~$N_f$ ($-0.95$, $-0.49$, and~$+0.17$
for~$N_f=8$, $32$, and~$64$, respectively), and crosses zero
at~$N_f=N_f^{\text{c}}\simeq56$ (by interpolation). Hence, a fixed point
appears only for~$N_f\gtrsim N_f^{\text{c}}$; for~$N_f=64$, it is located
at~$\kappa_*\simeq1.26$. For larger~$\kappa$, $\gamma_A>0$, and a zero can
arise only from a large negative fermion anomalous dimension. This happens
for~$N_f=1$ and~$2$, at~$\kappa\simeq6.1$ and~$6.7$, respectively; we discuss
it in~Sect.~\ref{sec:4.2}.

The value of~$N_f^{\text{c}}$ depends on the treatment of~$\mathcal{K}_\tau$. We
have repeated the above analysis with~$\mathcal{K}_\tau$ replaced by the
stationary solution of~Eq.~\eqref{eq:(2.18)}. In this case, the maximum
of~$\beta_\kappa/\kappa$ hardly grows with~$N_f$ ($-0.97$, $-0.88$, $-0.78$,
and~$-0.75$ for~$N_f=8$, $16$, $32$, and~$64$, respectively), and no zero is
found up to~$N_f=64$. Moreover, for~$N_f\gtrsim24$, the stationary solution
ceases to exist at~$\kappa\simeq1.0$--$1.6$, where $\gamma_A$ reaches~$-1$.
This is the resonance discussed in~Sect.~\ref{sec:3}: expanding the stationary
solution as~$\mathcal{K}(k^2)=1+K_1k^2+\dotsb$, Eq.~\eqref{eq:(2.18)} gives
$K_1=\mathcal{L}'(0)/[2(1+\gamma_A)]$ in~$D=4$, which diverges
as~$\gamma_A\to-1$, so that $\gamma_A$ cannot cross~$-1$ continuously.
Therefore, the fixed point found with~$\mathcal{K}_\tau=1$ is not robust
against the treatment of~$\mathcal{K}_\tau$, and we can only say that a fixed
point at weak Pauli coupling, if any, requires at least several tens of
flavors.

\subsection{Fixed point at a finite mass parameter}
\label{sec:4.2}
For~$N_f=1$, there is no fixed point at weak Pauli coupling. The mass
parameter, however, is not protected by the chiral symmetry, and a fixed point
need not lie on the branch of the critical mass connected to~$m=0$, which we
have followed so far. Indeed, the fixed point~$\mathcal{B}$
of~Ref.~\cite{Gies:2020xuh} has a finite mass parameter. We have thus searched
the whole $(\kappa,m)$ plane for the simultaneous zeros of~$\beta_\kappa$
and~$dm_\tau/d\tau$. In the subspace~$e_\tau=0$, the right-hand side of the
GFERG equation for our ansatz consists only of one-loop terms, and this search
involves no truncation of the loop expansion. The approximations lie in the
ansatz itself, including the choice of~$\mathcal{K}$, and in the numerical
interpolation in~$m$ described below. We have computed the one-loop
coefficients at~$m=-3$, $-2$, $-1.5$, $-1$, $-0.6$, $-0.3$, $0$, $0.3$, $0.6$,
$0.7$, $0.8$, and~$0.9$, and at~$m=1.5$, $2$, and~$3$, and interpolate them
in~$m$ by the Lagrange polynomial through the three nearest of these values.
The point~$m=1$ is special: the fermion propagator is~$h_F(0)=i/(1-m)$ at zero
momentum and has a pole there. We do not interpolate across this point, and
the interval~$0.95<m<1.5$ is not covered.

\begin{figure}[htbp]
\centering
\includegraphics[width=0.8\textwidth]{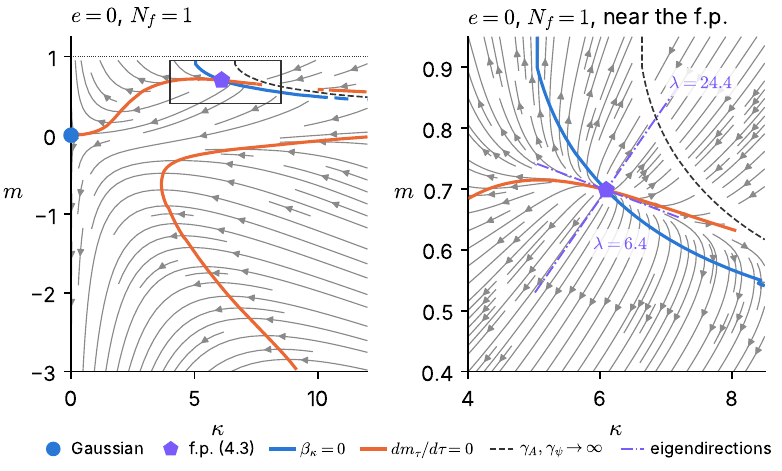}
\caption{RG flow in the $(\kappa,m)$ plane for~$e_\tau=0$, $N_f=1$, $D=4$,
and~$\mathcal{K}=1$; the flow is exactly of one-loop order. Left: the whole
plane. Right: the neighborhood of the fixed point~\eqref{eq:(4.3)} (the
rectangle in the left panel). The blue and orange curves are~$\beta_\kappa=0$
and~$dm_\tau/d\tau=0$, respectively, and the gray curves show the direction of
the flow toward the IR (their lengths and spacing do not represent the speed).
The pentagon is the fixed point~\eqref{eq:(4.3)}, and the circle is the
Gaussian fixed point. The dash-dotted lines in the right panel are the
eigendirections at the fixed point, with the eigenvalues indicated; both are
relevant. On the dashed curve, $\gamma_A$ and~$\gamma_\psi$ diverge. The dotted
line~$m=1$ in the left panel is the pole of the fermion propagator~$h_F$ at
zero momentum; the region~$1.5\leq m\leq3$ (not shown) contains no zero.}
\label{fig:4}
\end{figure}

Figure~\ref{fig:4} shows the curves~$\beta_\kappa=0$ and~$dm_\tau/d\tau=0$
for~$\mathcal{K}=1$. The curve~$dm_\tau/d\tau=0$ has two branches: the one
connected to~$m=0$, which is the critical-mass branch of~Fig.~\ref{fig:3} and
approaches~$m\simeq0.7$ for~$\kappa\gtrsim3$, and the one at negative~$m$
for~$\kappa\gtrsim3.6$, on which $\beta_\kappa/\kappa\simeq-2$ to~$-4$. The
curve~$\beta_\kappa=0$ exists only for~$\kappa\gtrsim5$
and~$0.5\lesssim m\lesssim0.95$. On the dashed curve, the linear system that
determines $\gamma_A$ and~$\gamma_\psi$ [Eq.~(3.3)
of~Ref.~\cite{Nagao:2026gbq}] becomes singular, and $\gamma_A$ and~$\gamma_\psi$
diverge. The sign changes of~$\beta_\kappa$ and~$dm_\tau/d\tau$ across this
curve are not zeros, and we exclude them. In the
region~$0<\kappa\leq12$ with~$-3\leq m\leq0.95$ or~$1.5\leq m\leq3$, we find a
single fixed point; for~$\mathcal{K}=1$, it is located at
\begin{equation}
   \kappa_*=6.09,\qquad m_*=0.698,\qquad
   \gamma_A^*=1.47,\qquad\gamma_\psi^*=-1.56,
\label{eq:(4.3)}
\end{equation}
which lies on the critical-mass branch connected to~$m=0$. With~$\mathcal{K}$
replaced by the stationary solution to~Eq.~\eqref{eq:(2.18)}, the zero
of~$\beta_\kappa$ along this branch moves to
\begin{equation}
   \kappa_*=8.61,\qquad m_*=0.673,\qquad
   \gamma_A^*=3.70,\qquad\gamma_\psi^*=-2.81.
\label{eq:(4.4)}
\end{equation}
The eigenvalues of the linearized flow in the $(\kappa,m)$ plane are~$24.4$
and~$6.4$ for~Eq.~\eqref{eq:(4.3)}, and~$42.9$ and~$11.3$
for~Eq.~\eqref{eq:(4.4)} (with~$\mathcal{K}$ held at its stationary form).
Both directions are relevant: the flow emanates from the fixed point toward
the IR (Fig.~\ref{fig:4}). In most of the plane, $\kappa_\tau$ decreases while
$m_\tau$ runs away from the critical branches. In the direction of the gauge
coupling, $\lambda_e=-\gamma_A^*=-1.47$ and~$-3.70$, respectively
[Eq.~\eqref{eq:(2.4)}], and the gauge coupling is irrelevant. The fixed point
thus has two relevant directions, those of~$\kappa$ and~$m$.

Unlike the large-$N_f$ Pauli fixed point, this fixed point is not caused by
the vacuum polarization, since~$\gamma_A^*>0$. It is caused by the large
negative fermion anomalous dimension: the term~$2\gamma_\psi$
overcompensates~$1+\gamma_A$ in~Eq.~\eqref{eq:(2.18)}. Since~$|\gamma_\psi|>1$,
the scaling dimension of the fermion deviates from its canonical value by more
than unity. The ordering of operators by their canonical dimensions, on which
our ansatz is based, is then no longer justified. The large difference
between~Eqs.~\eqref{eq:(4.3)} and~\eqref{eq:(4.4)} also shows that the
position of the fixed point is not under control. We regard its existence as
being outside the range of validity of our truncation. The irrelevance of the
gauge coupling, on the other hand, requires only~$\gamma_A^*>0$, which holds
with a large margin in both cases.

\subsection{Comparison with the functional RG}
\label{sec:4.3}
The fixed point~\eqref{eq:(4.3)} shares several features with the fixed
point~$\mathcal{B}$ of~Ref.~\cite{Gies:2020xuh}: a vanishing gauge coupling, a
strong Pauli coupling, a finite mass parameter, a large negative fermion
anomalous dimension, and two relevant directions (Table~\ref{table:4}). The
values of~$\kappa_*$ and~$m_*$ cannot be compared directly, since the
regularizations and the normalizations of the couplings are different. In our
case, the finite~$m_*$ is the value of the critical mass, $dm_\tau/d\tau=0$, and
the fixed point lies on the critical surface. In general, at a fixed point,
the dimensionful mass merely scales with the cutoff, so that a finite
dimensionless mass parameter does not imply a physical fermion mass. No fixed
point corresponding to the fixed point~$\mathcal{C}$
of~Ref.~\cite{Gies:2020xuh} is found: the critical surface is shifted
to~$m_c\propto\kappa^2$ by the chirality breaking of~$h_F$ and the Pauli term,
and the critical-mass branch has no zero of~$\beta_\kappa$
for~$\kappa\lesssim5$ (Fig.~\ref{fig:4}).
\begin{table}[htbp]
\centering\small
\begin{tabular}{lcccc}
\toprule
& $\mathcal{B}$~\cite{Gies:2020xuh} & $\mathcal{C}$~\cite{Gies:2020xuh}
& Eq.~\eqref{eq:(4.3)} & Eq.~\eqref{eq:(4.4)}\\
\midrule
$(\kappa_*,m_*)$ & $(5.09,0.328)$ & $(3.82,0)$ & $(6.09,0.698)$
& $(8.61,0.673)$\\
$\gamma_A^*$ ($=\eta_A/2$) & $0.27$ & $0$ & $1.47$ & $3.70$\\
$\gamma_\psi^*$ ($=\eta_\psi/2$) & $-0.69$ & $-0.50$ & $-1.56$ & $-2.81$\\
relevant directions & $2$ & $3$ & $2$ ($\kappa$, $m$) & $2$ ($\kappa$, $m$)\\
gauge coupling & relevant & relevant & irrelevant & irrelevant\\
\bottomrule
\end{tabular}
\caption{Fixed points with~$e_*=0$ for~$N_f=1$ and~$D=4$: the fixed
points~$\mathcal{B}$ and~$\mathcal{C}$ of the functional RG
analysis~\cite{Gies:2020xuh} and those found here in~Sect.~\ref{sec:4.2}
($\mathcal{K}=1$ and the stationary~$\mathcal{K}$, respectively). At the
latter, $\lambda_e=-\gamma_A^*$ [Eq.~\eqref{eq:(2.4)}]. The values
of~$\kappa_*$ and~$m_*$ depend on the regularization and the normalization and
cannot be compared directly.}
\label{table:4}
\end{table}

The essential difference lies in the direction of the gauge coupling.
In~Ref.~\cite{Gies:2020xuh}, the fixed points~$\mathcal{B}$ and~$\mathcal{C}$
are both IR repulsive in the direction of the gauge coupling, i.e., the gauge
coupling is relevant. The UV critical surface of~$\mathcal{B}$ then contains
trajectories with~$e\neq0$, and $\mathcal{B}$ defines a UV complete QED with a
nonzero charge, specified by two parameters. At the fixed
point~\eqref{eq:(4.3)}, on the other hand, the gauge coupling is irrelevant.
The UV critical surface is contained in the subspace~$e_\tau=0$, and all the
trajectories emanating from the fixed point stay there. A trajectory
with~$e_\tau\neq0$ moves away from the fixed point toward the UV. This fixed
point therefore cannot provide a UV completion of QED with a nonzero charge.

In the GFERG, a relevant gauge coupling at a fixed point with~$e_*=0$
requires~$\gamma_A^*<0$, i.e., a dominant vacuum polarization. The large-$N_f$
Pauli fixed point of~Sect.~\ref{sec:3} is an example, but at finite~$N_f$ it
requires at least several tens of flavors (Sect.~\ref{sec:4.1}).

\section{Full flow for $N_f=1$ in $D=4$}
\label{sec:5}
\subsection{Coefficients and numerical method}
\label{sec:5.1}
In this section, we consider the full flow for~$N_f=1$ in~$D=4$, in which the
gauge coupling is nonzero, on the critical surface connected to the Gaussian
fixed point (Sect.~\ref{sec:5.2}). We have evaluated the loop coefficients
$a$, $b$, $c_\gamma$, $c_\sigma$, and~$\mathcal{L}(k^2)$ in~Eq.~\eqref{eq:(2.17)}
up to three loops; only the three-loop term of~$c_\sigma$ could not be
determined reliably (Sect.~\ref{sec:5.4}). Table~\ref{table:5} lists the
coefficients at the Gaussian point. The multi-loop terms are computed
for~$N_f=1$ only.
\begin{table}[htbp]
\centering\small
\begin{tabular}{clcccc}
\toprule
loops & monomial & $a$ & $b$ & $c_\sigma$ & $\mathcal{L}(0)$\\
\midrule
1 & $e^2$ & $3.799\times10^{-2}$ & $5.241\times10^{-3}$ & & $1/6\pi^2$\\
1 & $e\kappa$ & $2.483\times10^{-2}$ & $9.421\times10^{-2}$ & & $-7.908\times10^{-2}$\\
1 & $\kappa^2$ & $6.105\times10^{-2}$ & $-4.087\times10^{-2}$ & & $-2.602\times10^{-2}$\\
1 & $e^3$ & & & $1.73\times10^{-4}$ & \\
1 & $e^2\kappa$ & & & $-1.27\times10^{-2}$ & \\
1 & $e\kappa^2$ & & & $2.36\times10^{-2}$ & \\
1 & $\kappa^3$ & & & $5.94\times10^{-3}$ & \\
\midrule
2 & $e^4$ & $4.68\times10^{-5}$ & $-1.780\times10^{-4}$ & & $-1.8\times10^{-4}$ $^\dagger$\\
2 & $e^3\kappa$ & $-3.95\times10^{-4}$ & $2.12\times10^{-4}$ & & $1.9\times10^{-4}$\\
2 & $e^2\kappa^2$ & $-7.3\times10^{-5}$ & $-6.4\times10^{-5}$ & & $7.0\times10^{-4}$\\
2 & $e\kappa^3$ & $1.77\times10^{-4}$ & $-7.0\times10^{-5}$ & & $4.3\times10^{-5}$\\
2 & $e^5$ & & & $1.04\times10^{-4}$ & \\
2 & $e^4\kappa$ & & & $1.74\times10^{-4}$ & \\
2 & $e^3\kappa^2$ & & & $-5.7\times10^{-5}$ & \\
2 & $e^2\kappa^3$ & & & $1.80\times10^{-4}$ & \\
2 & $e\kappa^4$ & & & $3.0\times10^{-5}$ & \\
\midrule
3 & & $\lesssim3\times10^{-6}$ & $\lesssim5\times10^{-6}$ & undetermined & not included\\
\bottomrule
\end{tabular}
\caption{The loop coefficients~$a$, $b$, $c_\sigma$, and~$\mathcal{L}(0)$
defined in~Eq.~\eqref{eq:(2.17)}, which enter the right-hand sides
of~Eqs.~\eqref{eq:(2.18)} and~\eqref{eq:(2.19)}, at~$m=0$, $\mathcal{K}=1$,
$D=4$, and~$N_f=1$. Since $\gamma_A$ and~$\gamma_\psi$ enter these coefficients
linearly and are determined algebraically (see below Eq.~\eqref{eq:(2.20)}),
we list only the parts independent of~$\gamma_A$ and~$\gamma_\psi$; the terms
proportional to~$\gamma_A$ and~$\gamma_\psi$ contribute at the next order in
the couplings. $^\dagger$This entry, the $e^4$ term of~$\mathcal{L}(0)$ in QED
without the Pauli term, depends on the resolution of the numerical
integration. The same term was computed in~Ref.~\cite{Nagao:2026gbq} with a
finer resolution, from the expression for the photon two-point function
in~Eq.~(4.3) of~Ref.~\cite{Nagao:2026vaf}, as~$-1.302\times10^{-4}$. Replacing
our value by it changes~$\gamma_A$ by only~$1.2\%$ at~$e^2=4$. The three-loop
entries are quasi-Monte Carlo estimates; see~Sect.~\ref{sec:5.4}.}
\label{table:5}
\end{table}
The one-loop QED values reproduce~$a=3/8\pi^2$~\cite{Nagao:2026gbq},
$b=0.005242$~\cite{Nagao:2026vaf}, and~$\mathcal{L}(0)=1/6\pi^2$, i.e.,
$\gamma_A=e^2/12\pi^2$, which is the universal one-loop value of the photon
anomalous dimension~\cite{Nagao:2026vaf,Nagao:2026gbq}. The
coefficient~$c_\sigma$ has a pure-gauge term~$\propto e^3$ already at one loop,
although~$m=0$, owing to the chirality breaking of~$h_F$ (Sect.~\ref{sec:2});
it gives the induced Pauli coupling discussed in~Sect.~\ref{sec:5.3}.

The loop coefficients are evaluated by the source-inversion method of
Appendix~A of~Ref.~\cite{Nagao:2026uxd}, with the following essential
differences. First, in Ref.~\cite{Nagao:2026uxd}, the four-fermion interactions
do not contribute to the photon two-point function, which was therefore
evaluated from~Eq.~(4.3) of~Ref.~\cite{Nagao:2026vaf}, and the background
fields of the source inversion consisted of fermion fields only. Here, the
Pauli term does contribute to the photon two-point function, and the
three-point function has an external photon. We therefore include background
photons in the source inversion; they are represented by commuting nilpotent
parameters, so that the series of the source inversion still terminates. The
Pauli vertex is constructed from the same $-1$~composite fields as the gauge
vertex, with~$\gamma_\nu$ replaced by~$2\kappa_\tau\sigma_{\nu\alpha}k_\alpha$. The
three-point function is projected by~Eq.~\eqref{eq:(2.21)} at~$|k|=0.02$; the
contribution of its $O(k^0)$~part to this projection, which is odd in~$k$, is
removed by averaging the projected value over~$\pm k$. The multi-loop part of
the photon two-point function is obtained from the coefficient of the two
background photons~$\mathcal{A}(\pm k)$: denoting its transverse part
by~$\Pi(k)$, we set $\mathcal{L}(k^2)=-e^{-2k^2}[\Pi(k)-\Pi(0)]/k^2$, where the
factor~$e^{-2k^2}/k^2$ converts~$\Pi$ to the normalization of~$\mathcal{K}_\tau$
in the $-1$~variables. Since $\Pi(0)$, which vanishes by gauge invariance, does
not vanish exactly in the numerical integration, we subtract it before dividing
by~$k^2$.

Second, the two-loop integrals are evaluated by Gauss--Legendre quadrature as
in~Ref.~\cite{Nagao:2026uxd}. Unlike there, the photon momentum and
polarization single out directions in space, and the integrand depends on
their orientation relative to the loop momenta. We average over this
orientation exactly, by using that the projected quantities are polynomials of
low degree in the unit vectors along the photon momentum and polarization: the
average over all orientations is then reproduced by that over a finite set of
directions, namely, the $D(D-1)$ pairs of different coordinate axes
for~$c_\sigma$, and the $24$~vertices of the 24-cell, a spherical
$5$-design~\cite{Delsarte:1977}, for the photon two-point function. A
stochastic average with random orientations instead gives errors of
order~$100\%$, since the integrand is large with both signs (the dash-dotted
curves in~Fig.~\ref{fig:5}). The multi-loop tables are made
at~$m=0$; along the flows below, $|m_\tau|\lesssim0.08$.

\subsection{Flow}
\label{sec:5.2}
For each initial condition~$(e_0^2,\kappa_0)$ with~$\mathcal{K}_0=1$, we tune
the initial mass by bisection so that $m_\tau$ does not run away, as
in~Ref.~\cite{Nagao:2026uxd}. All the two- and three-loop terms are included,
except for the undetermined three-loop term of~$c_\sigma$ (Sect.~\ref{sec:5.4}).
Since the precision of the tuning is exhausted at~$\tau\simeq7.5$, beyond
which $m_\tau$ starts to deviate from the critical value, we show the flows up
to~$\tau=7.5$.

The bisection selects the critical surface connected to the Gaussian
point~$m=0$, and our flows explore only its part at weak Pauli coupling.
Unlike in~Sect.~\ref{sec:4.2}, where the whole $(\kappa,m)$ plane was searched
for~$e_\tau=0$, we do not examine here, for~$e_\tau\neq0$, the branch at
negative~$m$ or the strong-coupling part of the branch connected to~$m=0$,
near~$m\simeq0.7$ in~Fig.~\ref{fig:4}, on which the fixed
point~\eqref{eq:(4.3)} lies. The multi-loop tables are made at~$m=0$ only.
This suffices along our flows, on which~$|m_\tau|\lesssim0.08$, but not in
those regions; a search for fixed points with~$e_\tau\neq0$ there requires the
multi-loop coefficients as functions of~$m$. As a rough check, we have
computed the two-loop tables also at~$m=-0.5$ for~$\kappa=0$. Interpolating
linearly between~$m=0$ and~$-0.5$, we find~$\gamma_A>0$ for~$-0.5\leq m\leq0$
and~$e^2\leq60$; no fixed point with~$e_\tau\neq0$ appears in this range.

\begin{figure}[htbp]
\centering
\includegraphics[width=0.84\textwidth]{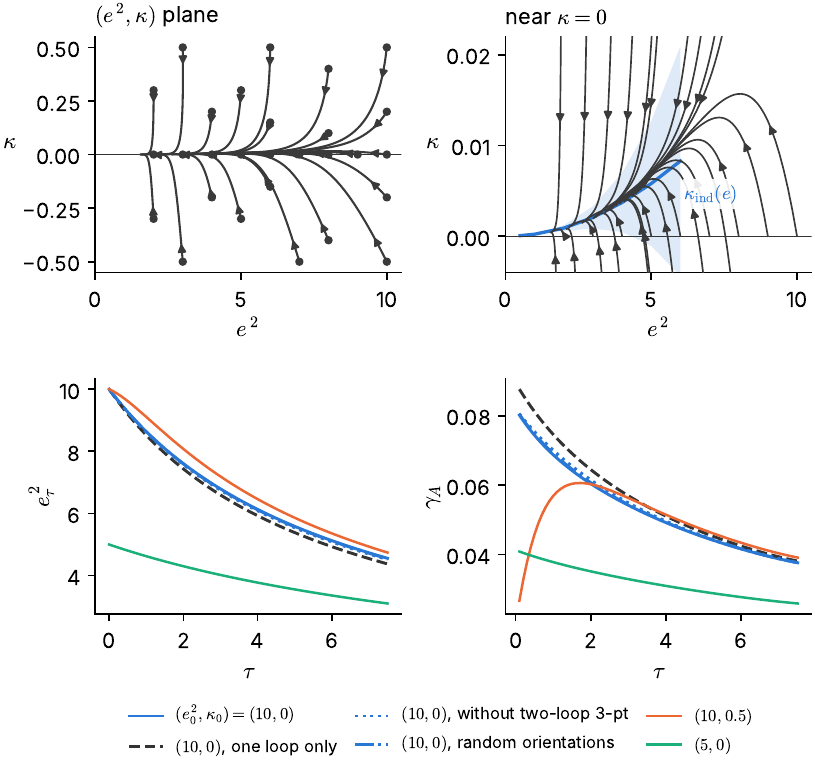}
\caption{Mass-tuned flows for~$D=4$, $N_f=1$, $\xi_\tau=0$,
and~$\mathcal{K}_0=1$, with the two- and three-loop terms except for the
undetermined three-loop term of~$c_\sigma$. Top left: trajectories in
the~$(e^2,\kappa)$ plane for~$28$ initial conditions (dots). Top right: the
same near~$\kappa=0$, together with the induced value~$\kappa_{\text{ind}}(e)$
and its uncertainty from the three-loop term (Fig.~\ref{fig:7}). Bottom:
$e_\tau^2$ and~$\gamma_A$ as functions of~$\tau$; for~$(e_0^2,\kappa_0)=(10,0)$,
the results with the one-loop terms only (dashed), without the two-loop terms
of the three-point function (dotted), and with the two-loop coefficients
obtained by random orientations instead of the exact orientation average
(dash-dotted; see Sect.~\ref{sec:5.1}) are also shown.}
\label{fig:5}
\end{figure}

\begin{figure}[htbp]
\centering
\includegraphics[width=0.46\textwidth]{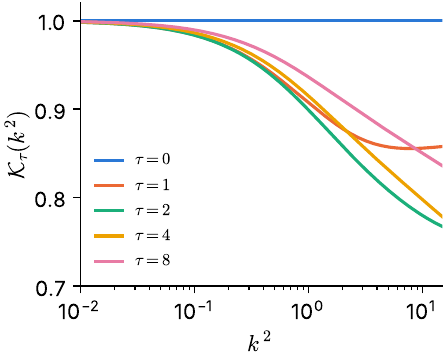}
\caption{$\mathcal{K}_\tau(k^2)$ along the flow from~$(e_0^2,\kappa_0)=(10,0)$
in~Fig.~\ref{fig:5} (the blue solid curves in the bottom panels).}
\label{fig:6}
\end{figure}

Figure~\ref{fig:5} shows the resulting flows in the~$(e^2,\kappa)$ plane
for~$28$ initial conditions. The Pauli coupling decays as~$e^{-\tau}$ even
from~$|\kappa_0|=0.5$, and, within a few units of~$\tau$, all the trajectories
are attracted to a common curve close to the induced
value~$\kappa=\kappa_{\text{ind}}(e)$ discussed in~Sect.~\ref{sec:5.3}, on which
$\kappa_\tau\sim10^{-3}$--$10^{-2}$ is determined by~$e_\tau$. This is consistent
with the result of~Sect.~\ref{sec:4} that, for~$N_f=1$, the Pauli coupling is
irrelevant at weak coupling. At~$\tau=6$,
$(e^2,\kappa)=(5.127,7.29\times10^{-3})$, $(3.361,2.91\times10^{-3})$,
and~$(5.360,8.16\times10^{-3})$ for~$(e_0^2,\kappa_0)=(10,0)$, $(5,0)$,
and~$(10,0.5)$, respectively. The photon anomalous dimension~$\gamma_A$ stays
positive along the flow; thus there is no fixed point with~$e_\tau\neq0$, and
$e_\tau^2$ is marginally irrelevant. The multi-loop terms reduce~$\gamma_A$
by~$4\%$ at~$e^2=4$ and by~$10\%$ at~$e^2=10$; with the one-loop terms only, we
have~$e_\tau^2=4.93$ at~$\tau=6$ for~$e_0^2=10$. The flows of~$e_\tau$, $m_\tau$,
and~$\mathcal{K}_\tau$ are practically independent of the Pauli sector; for
instance, removing the two-loop terms of the three-point function
changes~$e^2(\tau=6)$ only by~$0.8\%$. Hence, on the critical surface
connected to~$m=0$ and in the region we studied, the only fixed point
for~$D=4$ and~$N_f=1$ is the Gaussian one, with~$\lambda_m=1$,
$\lambda_\kappa=-1$, $\lambda_\xi=0$, and a marginally irrelevant~$e_\tau$.
Figure~\ref{fig:6} shows~$\mathcal{K}_\tau(k^2)$, which develops a logarithmic
decrease for~$k^2\gtrsim1$ as in QED without the Pauli
term~\cite{Nagao:2026gbq}.

\subsection{Induced Pauli coupling}
\label{sec:5.3}
In~Eq.~\eqref{eq:(2.18)}, $c_\sigma$ contains terms with no power
of~$\kappa_\tau$ ($e^3$ at one loop, $e^5$ at two loops, and~$e^7$ at three
loops), so that $\kappa_\tau=0$ is not preserved, and the Pauli term is
necessarily generated if~$e_\tau\neq0$. These terms, however, do not affect the
irrelevance of~$\kappa_\tau$. Whether $\kappa_\tau$ is relevant or irrelevant is
determined by the coefficient of the term linear in~$\kappa_\tau$
in~$d\kappa_\tau/d\tau$, i.e., $-(1+\gamma_A+2\gamma_\psi)$ together with the
contributions of the terms in~$c_\sigma$ linear in~$\kappa_\tau$; this
coefficient is close to or below~$-1$ for~$e^2\lesssim10$, since $2\gamma_\psi$
and~$\gamma_A$ are positive there. The $\kappa_\tau$-independent terms
of~$c_\sigma$ act as an inhomogeneous source, which determines only the value to
which $\kappa_\tau$ is attracted. In~$D=4$, $e_\tau$ runs only logarithmically,
whereas $\kappa_\tau$ relaxes as~$e^{-\tau}$. Therefore, within~$\tau\sim1$,
$\kappa_\tau$ catches up with the value at which~$d\kappa_\tau/d\tau\simeq0$ for
the current~$e_\tau$,
\begin{equation}
   \kappa_{\text{ind}}(e)\simeq
   \frac{c_\sigma(e,\kappa=0,m_c)}{2(1+\gamma_A+2\gamma_\psi)},
\label{eq:(5.1)}
\end{equation}
and then follows~$e_\tau$. In the theory space, the trajectories are thus
attracted to a one-dimensional manifold, which is approximately given
by~$\kappa=\kappa_{\text{ind}}(e)$: at low energies, the Pauli coupling is not an
independent parameter but is determined by~$e$, as irrelevant couplings are in
general determined by the relevant and marginal ones~\cite{Polchinski:1983gv}.
As remarked in~Sect.~\ref{sec:2}, the induced value is nonuniversal, like the
critical mass.

For each value of~$e$, we define~$\kappa_{\text{ind}}(e)$ as the solution
of~$d\kappa_\tau/d\tau=0$, with $e$ held fixed, $m$ set to the critical
value~$m_c$, and~$\mathcal{K}=1$; Eq.~\eqref{eq:(5.1)} is its approximate
form. Figure~\ref{fig:7} shows~$\kappa_{\text{ind}}/e^3$. Up to two loops, we
obtain~$\kappa_{\text{ind}}/e^3=2.13$, $3.14$, $4.64$, and~$5.64\times10^{-4}$
for~$e^2=1$, $2$, $4$, and~$6$, respectively. If the two-loop terms
of~$c_\sigma$ are omitted (dotted curve in~Fig.~\ref{fig:7}), the value
at~$e^2=4$ decreases by about~$40\%$; the two-loop terms are thus not small
for~$e^2\gtrsim4$. The two-loop terms themselves are accurate to~$3$--$8\%$,
as indicated by the residual of the WT identity~$c_\gamma=ae_\tau$
in~Eq.~\eqref{eq:(2.19)}, so that the numerical uncertainty
of~$\kappa_{\text{ind}}$ is only about~$3\%$. The uncertainty due to the
undetermined three-loop term is much larger, as discussed
in~Sect.~\ref{sec:5.4}.

The actual flow approaches~$\kappa_{\text{ind}}$ but does not coincide with it.
For~$\tau\gtrsim5$, $\kappa_\tau$ lies $10$--$20\%$
above~$\kappa_{\text{ind}}(e_\tau)$ (thin curves in~Fig.~\ref{fig:7}; see also the
top right panel of~Fig.~\ref{fig:5}). This is because $\kappa_\tau$ lags behind
the slowly decreasing~$e_\tau$, and because $\mathcal{K}_\tau$ deviates from
unity along the flow.

\subsection{The three-loop term of $c_\sigma$}
\label{sec:5.4}
\begin{figure}[htbp]
\centering
\includegraphics[width=0.54\textwidth]{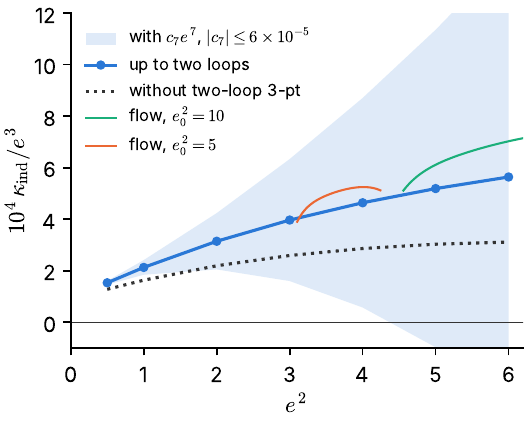}
\caption{$\kappa_{\text{ind}}/e^3$ as a function of~$e^2$ ($\mathcal{K}=1$
and~$m=m_c$). The band shows the effect of a three-loop term~$c_7e^7$
in~$c_\sigma$ with~$|c_7|\leq6\times10^{-5}$, and the dotted curve is obtained
without the two-loop terms of the three-point function. The thin curves
are~$\kappa_\tau/e_\tau^3$ along the flows from~$(e_0^2,\kappa_0)=(10,0)$
and~$(5,0)$ in~Fig.~\ref{fig:5} ($2<\tau<7.5$).}
\label{fig:7}
\end{figure}

The three-loop~$e^7$ term of~$c_\sigma$ could not be determined reliably. Its
integrand is concentrated at small loop momenta, $|q|\sim0.2$--$0.5$, where it
is larger by~$10^5$--$10^6$ than in the bulk of the sampling region of the
Gaussian weight, and the quasi-Monte Carlo estimates obtained with
independently scrambled Sobol sequences~\cite{Sobol:1967,Owen:1997} (the
orientation of the loop momenta relative to the photon momentum and
polarization being sampled together with the other variables) differ by an
order of magnitude; a single sample out of~$64$ can carry~$80\%$ of the total.
A deterministic six-dimensional quadrature would be required, which is beyond
our present computational resources. We therefore exclude this term from the
central values, and estimate its natural size from the ratio of the two- and
one-loop coefficients as~$|c_7|\lesssim6\times10^{-5}$. The resulting band
in~Fig.~\ref{fig:7} shows that $\kappa_{\text{ind}}$ is determined perturbatively
only for~$e^2\lesssim1$--$2$; for~$e^2\gtrsim4$, it is uncertain by a factor
of order unity, and for~$e^2=6$, even its sign depends on the three-loop term.
The qualitative conclusions, i.e., that $\kappa_\tau$ is irrelevant and is
slaved to~$e_\tau$ and that $e_\tau$ is marginally irrelevant, are not affected:
since the $e^7$ term does not depend on~$\kappa_\tau$, it only
shifts~$\kappa_{\text{ind}}$ and does not change the coefficient that determines
the irrelevance of~$\kappa_\tau$ (see~Sect.~\ref{sec:5.3}). Quantitatively,
varying~$c_7$ within~$|c_7|\leq6\times10^{-5}$ changes~$e^2(\tau=6)$ by
about~$\pm2\%$ for~$(e_0^2,\kappa_0)=(10,0)$.

\section{Conclusion}
\label{sec:6}
In this paper, we have studied the GFERG flow of QED with the Pauli term.
Since the Pauli term written in terms of the $-1$~variables is gauge invariant
by itself, the structure of the flow implied by the gauge WT identity is
unchanged, and the gauge coupling flows only through~$\gamma_A$ according
to~Eq.~\eqref{eq:(2.3)}. The flow of the Pauli coupling~$\kappa_\tau$ requires
the magnetic part of the fermion--photon three-point function. We have
numerically evaluated it, together with all the other terms of the GFERG
equation up to three loops (the three-loop term of~$c_\sigma$ remains
undetermined), by extending the source-inversion method
of~Ref.~\cite{Nagao:2026uxd}.

A direct consequence of the exact WT identity is that, at any fixed point
with~$e_*=0$, the eigenvalue in the direction of the gauge coupling
is~$\lambda_e=-\gamma_A^*$ [Eq.~\eqref{eq:(2.4)}]. The gauge coupling is
relevant at such a fixed point if and only if the vacuum polarization
makes~$\gamma_A^*$ negative, and the fermion anomalous dimension, however
large, does not contribute. This statement does not depend on the truncation
of the fermion sector.

Since all the two- and three-loop terms of the GFERG equation are proportional
to the gauge coupling, the subspace~$e_\tau=0$, in which the fixed points
of~Refs.~\cite{Gies:2020xuh,Gies:2022aiz,Gies:2024ugz} are located, is
described by one-loop terms alone, without truncation of the loop expansion,
for any~$N_f$. In the leading large-$N_f$ approximation, the vacuum
polarization induced by the Pauli coupling gives a fixed point with~$e_*=0$,
$\gamma_A^*=-1$, and three relevant directions, including the gauge coupling.
Every trajectory with~$\Tilde{\kappa}\neq0$ in
the~$(\Tilde{e},\Tilde{\kappa})$ plane emanates from this fixed point, and
those on one side of a separatrix approach the Gaussian fixed point in the IR.
In lower dimensions, there is no such fixed point in the same approximation:
in~$D=3$, the Pauli coupling is irrelevant at both the Gaussian and the IR
fixed points, since~$C_2(0)>0$, and in~$D=2$, it is marginally irrelevant at
the Gaussian fixed point, in accordance with the large-$N$ beta function of
the equivalent four-fermion interaction in the repulsive pseudoscalar channel.
The $N_f$-independent effects on the fermion line, i.e., the fermion anomalous
dimension and the additive mass renormalization proportional to~$\kappa^2$,
eliminate the fixed point at weak Pauli coupling unless $N_f$ is at least
several tens: with~$\mathcal{K}_\tau=1$, a fixed point appears only
for~$N_f\gtrsim56$, and with the stationary~$\mathcal{K}_\tau$, none is found
up to~$N_f=64$. For~$N_f\leq4$, this conclusion does not depend on the
truncation as far as the weak Pauli coupling is concerned, since the
coefficient of the leading correction to~$\beta_\kappa$ is determined by
one-loop terms alone [Eq.~\eqref{eq:(4.2)}].

For~$N_f=1$, the search over the whole $(\kappa,m)$ plane in the
subspace~$e_\tau=0$ finds a single fixed point, at strong Pauli coupling and a
finite mass parameter [Eqs.~\eqref{eq:(4.3)} and~\eqref{eq:(4.4)}], which is
the value of the critical mass there. It resembles the fixed
point~$\mathcal{B}$ of the asymptotically safe QED
of~Ref.~\cite{Gies:2020xuh}, which also has a finite mass parameter: the
fermion anomalous dimension is large and negative, and there are two relevant
directions. Our conclusion, however, differs from that
of~Ref.~\cite{Gies:2020xuh} in an essential point. At our fixed point,
$\gamma_A^*>0$, so that the gauge coupling is irrelevant, and the two relevant
directions are those of the Pauli coupling and the mass. The trajectories
emanating from the fixed point stay in the subspace~$e_\tau=0$, and the fixed
point thus cannot provide a UV completion of QED with a nonzero charge. No
counterpart of the fixed point~$\mathcal{C}$ in~Ref.~\cite{Gies:2020xuh} is
found in our analysis. For~$N_f=1$, the full flow on the critical surface
connected to~$m=0$ shows that the Pauli coupling is irrelevant and is
attracted to the induced value~$\kappa_{\text{ind}}(e_\tau)$, while the gauge
coupling is marginally irrelevant. Within our truncation, the Pauli term
therefore does not provide a UV completion of QED with~$N_f=1$, at least on the
critical surface connected to~$m=0$.

Several limitations of the present analysis should be mentioned. First, at the
fixed point found for~$N_f=1$, the fermion anomalous dimension exceeds unity
in magnitude, and its existence and position lie outside the range of validity
of our ansatz, which does not contain, e.g., the four-fermion operators
generated by the Pauli coupling~\cite{Gies:2024ugz} or the momentum dependence
of the vertices; the irrelevance of the gauge coupling there, which requires
only~$\gamma_A^*>0$, is more robust. Second, for~$e_\tau\neq0$
(Sect.~\ref{sec:5}), we have examined only the part of the critical surface
near~$m=0$, i.e., at weak Pauli coupling, since the multi-loop coefficients
are available only at~$m=0$ (apart from a rough check at~$m=-0.5$); a fixed
point with~$e_\tau\neq0$ elsewhere on the critical surface, e.g., near the
fixed point~\eqref{eq:(4.3)} at~$m\simeq0.7$, has not been excluded. Third,
the existence of the large-$N_f$ Pauli fixed point relies on the sign
of~$C_2(0)$. Since the Pauli coupling has mass dimension~$-1$, the $\kappa^2$
contribution to the vacuum polarization is power-like and its sign may depend
on the scheme, although the same sign is found in the functional
RG~\cite{Gies:2020xuh}. Moreover, the resonance of the $\mathcal{K}_\tau$ mode
at~$\gamma_A=-1$ makes it a fixed point of the couplings only. Fourth, the
three-loop term of~$c_\sigma$ is undetermined, which limits the quantitative
determination of~$\kappa_{\text{ind}}$ to~$e^2\lesssim2$; in any case,
$\kappa_{\text{ind}}$ is nonuniversal, like the critical mass. Finally, we have
worked in the Landau gauge; with our ansatz, the coefficients of the flow
equations depend on~$\xi_\tau$, as noted
in~Refs.~\cite{Nagao:2026vaf,Nagao:2026uxd}. The
relation~$\lambda_e=-\gamma_A^*$, however, holds in any gauge.

\section*{Acknowledgments}
We would like to thank Hidenori Sonoda for many years of valuable discussions.
The work of H.S. was supported by JSPS KAKENHI Grant Number JP23739313 and by
the MEXT program ``Supporting Pioneering Research through AI for 1,000
Discovery Challenges (SPReAD),'' Grant Number JPMXP1726257479. It was also
partially supported by the Quantum and Spacetime Research Institute, Kyushu
University.
We gratefully acknowledge substantial assistance from Claude (Anthropic) and
ChatGPT (OpenAI) in developing the numerical methods and the codes implementing
them, in carrying out the numerical computations and their cross-checks, and
in providing suggestions on the manuscript. Needless to say, any remaining
errors are our own.

\end{document}